\UseRawInputEncoding
\documentclass[nofootinbib,pra,twocolumn,showpacs,superscriptaddress,notitlepage,superscriptaddress,amsmath]{revtex4-2}
\usepackage[colorlinks=true,citecolor=blue,linkcolor=blue,urlcolor=blue]{hyperref}
\usepackage{xcolor}
\usepackage{amsmath,amssymb,bm,mathtools,amsthm, amssymb}
\usepackage{tikz}
\usepackage{physics}
\usepackage[compat=0.6]{yquant}
\usepackage{dsfont}
\usepackage{tabularx}
\usepackage[normalem]{ulem}
\usepackage[T1]{fontenc}

\def\ZZ{\mathbb Z}
\def\ebar{\mkern 1.5mu\overline{\mkern-1.5mu e \mkern-1.5mu}\mkern 1.5mu}
\def\mbar{\mkern 1.5mu\overline{\mkern-1.5mu m \mkern-1.5mu}\mkern 1.5mu}
\def\vac {{\bf{1}}}
\def\Xent {X_{\textrm{ent}}}
\def\Zent {Z_{\textrm{ent}}}

\newcommand{\cbox}[2]{\vcenter{\hbox{\includegraphics[width=#1em]{#2}}}}

\begin{document}

\begin{abstract}
The development of programmable quantum devices can be measured by the complexity of many-body states that they are able to prepare. Among the most significant are topologically ordered states of matter, which enable robust quantum information storage and processing. While topological orders are more readily accessible with qudits, experimental realisations have thus far been limited to lattice models of qubits.
Here, we prepare a ground state of the $\mathbb{Z}_3$ toric code state on 24 qutrits in a trapped ion quantum processor with fidelity per qutrit exceeding $96.5(3)\%$. We manipulate two types of defects which go beyond the conventional qubit toric code: a parafermion, and its bound state which is related to charge conjugation symmetry. We further demonstrate defect fusion and the transfer of entanglement between anyons and defects, which we use to control topological qutrits. Our work opens up the space of long-range entangled states with qudit degrees of  freedom for use in quantum simulation and universal error-correcting codes.
\end{abstract}
\title{Qutrit Toric Code and Parafermions in Trapped Ions}

\author{Mohsin Iqbal}
\thanks{These authors contributed equally.}
\affiliation{Quantinuum, Leopoldstrasse 180, 80804 Munich, Germany}
\author{Anasuya Lyons}
\thanks{These authors contributed equally.}
\affiliation{Department of Physics, Harvard University, Cambridge, MA 02138, USA}
\author{Chiu Fan Bowen Lo}
\affiliation{Department of Physics, Harvard University, Cambridge, MA 02138, USA}
\author{Nathanan Tantivasadakarn}
\affiliation{Walter Burke Institute for Theoretical Physics and Department of Physics, California Institute of Technology, Pasadena, CA 91125, USA}

\author{Joan Dreiling}
\affiliation{Quantinuum, 303 S Technology Ct, Broomfield, CO 80021, USA}

\author{Cameron Foltz}
\affiliation{Quantinuum, 303 S Technology Ct, Broomfield, CO 80021, USA}

\author{Thomas M. Gatterman}
\affiliation{Quantinuum, 303 S Technology Ct, Broomfield, CO 80021, USA}

\author{Dan Gresh}
\affiliation{Quantinuum, 303 S Technology Ct, Broomfield, CO 80021, USA}

\author{Nathan Hewitt}
\affiliation{Quantinuum, 303 S Technology Ct, Broomfield, CO 80021, USA}

\author{Craig A. Holliman}
\affiliation{Quantinuum, 303 S Technology Ct, Broomfield, CO 80021, USA}

\author{Jacob Johansen}
\affiliation{Quantinuum, 303 S Technology Ct, Broomfield, CO 80021, USA}

\author{Brian Neyenhuis}
\affiliation{Quantinuum, 303 S Technology Ct, Broomfield, CO 80021, USA}

\author{Yohei Matsuoka}
\affiliation{Quantinuum, 303 S Technology Ct, Broomfield, CO 80021, USA}

\author{Michael Mills}
\affiliation{Quantinuum, 303 S Technology Ct, Broomfield, CO 80021, USA}

\author{Steven A. Moses}
\affiliation{Quantinuum, 303 S Technology Ct, Broomfield, CO 80021, USA}

\author{Peter Siegfried}
\affiliation{Quantinuum, 303 S Technology Ct, Broomfield, CO 80021, USA}

\author{Ashvin Vishwanath}
\affiliation{Department of Physics, Harvard University, Cambridge, MA 02138, USA}
\author{Ruben Verresen}
\affiliation{Pritzker School of Molecular Engineering, University of Chicago, Chicago, IL 60637, USA}
\affiliation{Department of Physics, Harvard University, Cambridge, MA 02138, USA}
\author{Henrik Dreyer}
\email{henrik.dreyer@quantinuum.com}
\affiliation{Quantinuum, Leopoldstrasse 180, 80804 Munich, Germany}

\date{\today}

\maketitle

\section{Introduction}
The unprecedented tunability of quantum processors has opened up the \emph{on-demand} preparation and control of topologically ordered (TO) quantum states~\cite{semeghini_probing_2021,satzinger_realizing_2021, bluvstein_quantum_2022, iqbal_topological_2024, foss2023experimental, bluvstein_logical_2024, acharya_quantum_2024, reichardt_demonstration_2024, berthusen_experiments_2024}. Creating these states in programmable quantum computers does not only allow for the simulation of complex many-body systems, but also provides a code-space for quantum computation.
In particular, the quasiparticle excitations of TO phases of matter---known as anyons---exhibit exchange statistics beyond those familiar from bosons or fermions \cite{leinaas_theory_1977, goldin_representations_1980, wilczek_quantum_1982, frohlich_braid_1990}. The robust braiding of such anyons constitutes the primitive of topological quantum computation \cite{freedman_modular_2000, kitaev2003fault}.
This experimental program is being extended to increasingly complex TOs, including non-Abelian TO~\cite{iqbal2024non,xu2024non,minev2024realizing} as well as defects that enrich the computational power of Abelian TO~\cite{google2023non,xu2023digital}.

While anyons are pointlike deformations of the state, defects are associated to extended objects, like lattice dislocations or vortices in superfluids. As such, using defects to process quantum information in a topologically protected way is subject to more caveats than is the case of genuine anyonic excitations, as these rigid objects are harder to move and their braiding properties are more restrictive.
On the other hand, defects can exist in comparatively less exotic states: A familiar example is the non-Abelian Majorana defect which can be inserted into the toric code \cite{Bombin10}, whereas Majorana anyons require a non-Abelian topologically ordered state \cite{kitaev2006anyons}. In fact, defects can be seen as precursors to anyons. More precisely, defects are often related to global physical symmetries of the system and gauging these symmetries promotes defects into genuine anyons of a larger TO~\cite{Barkeshligauging19}. Thus, gauging introduces a hierarchy on the space of TOs~\cite{tantivasadakarn_hierarchy_2023}.

\begin{figure}
    \centering
\includegraphics[width=0.99\linewidth]{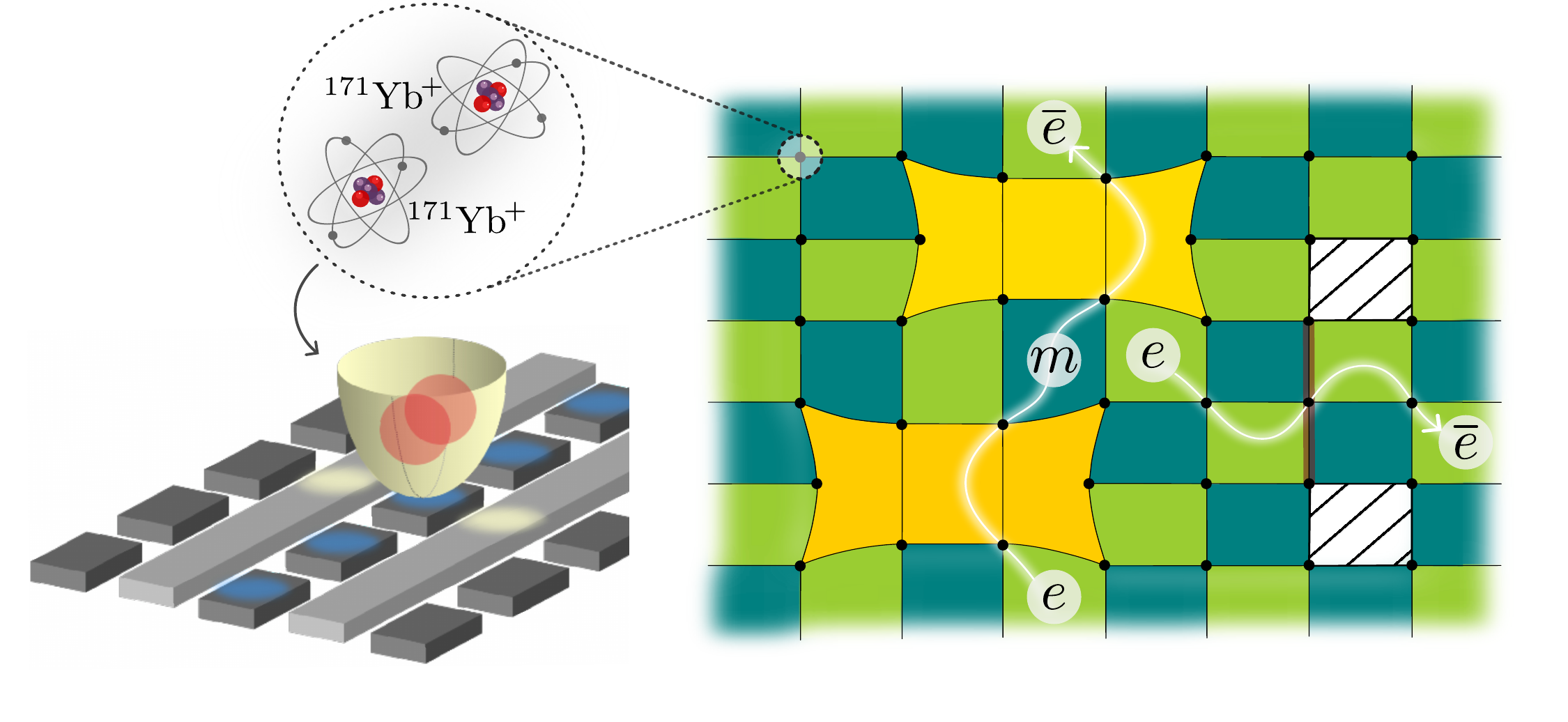}
    \caption{\textbf{Concept.} We start with 56 trapped $^{171}{\rm Yb}^{+}$ ions in a quantum charge-coupled device and algorithmically encode pairs of qubits into qutrits. This allows us to prepare ground states of $\mathbb{Z}_3$ toric codes on tori of up to $6\times 4$ qutrits. We conduct experiments to study the relationship between the anyons and the topological defects of this system, namely parafermion (shaded yellow) and charge conjugation defects (hatch pattern). \label{fig:concept}}
\end{figure}

For the conventional toric code \cite{kitaev2003fault}, the only defects are the aforementioned Majoranas, associated to the $e \leftrightarrow m$ symmetry of the toric code. The hierarchy accessible through gauging is thus restricted. A richer class opens up by upgrading qubits to qutrits: a toric code based on the gauge group $\mathbb{Z}_3$ hosts two kinds of charges and fluxes; e.g., the `electric' $e$-anyon is no longer its own anti-particle, the latter now denoted as $\ebar$, and similarly for the flux anyons $m,\mbar$. Correspondingly, a new ``charge conjugation'' symmetry emerges: $e \leftrightarrow \ebar$, $m \leftrightarrow {\mbar}$.
The resulting symmetry-enriched physics has received much attention, not least because it forms the backbone of certain proposals for obtaining universal non-Abelian error correction codes which can be prepared with constant-depth adaptive circuits~\cite{mochon_anyon_2004,Cui2015,Shawn15,Claire15, cong_universal_2017, giudice_trimer_2022, verresen_efficiently_2022,sptmeasure,bravyi2022adaptiveconstantdepthcircuitsmanipulating,tantivasadakarn_hierarchy_2023,PhysRevB.108.115144}.
Despite the theoretical interest, qudit-based TOs such as the $\mathbb{Z}_3$ toric code have so far eluded experimental observation both by devices with native qutrit degrees of freedom~\cite{edmunds_constructing_2024}, as well as qubit-based platforms.

\begin{figure*}[t]
	\centering    \includegraphics[width=0.84\linewidth]{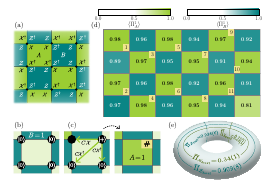}
	\caption{\textbf{Preparation of Qutrit Toric Code.} (a) Square lattice on a torus with qutrits on the vertices.
	(b) Qutrits are initialized in the $\ket{0}$ state, satisfying $B = +1$ (visually represented by the intense turquoise color of type-B plaquettes, in contrast to the faded green color of type-A plaquettes, which do not satisfy the $A=+1$ condition at this stage). (c)~Preparation of one of the type-A plaquettes. One of the qutrits is initialized in the $\ket{+}$ state. This qutrit is used as a control, and we apply C$\mathcal{X}$ or C$\mathcal{X}^\dagger$ gates to other qutrits in the plaquette. This leads to satisfying $A=+1$, indicated by a bright green color on the right-hand side of the arrow. A small square in one corner of the plaquette indicates the control qutrit. The number within the square denotes the order in which the corresponding stabilizer is prepared. (d)~Expectation values of projectors $\Pi_{\bullet}^1$ obtained by measuring qutrits in the $\mathcal{X}$ and $\mathcal{Z}$ basis. The maximum error in estimating the expectation values is $0.022$. The mean energy density, $\langle H \rangle/24 \geq -1$, is found to be $-0.945(3)$. (e)~Mean expectation values of projectors for the logical $\mathcal{X}$ and $\mathcal{Z}$ operators in two directions on the torus closely match the theoretical predictions: $\langle\Pi^1_{\mathcal{Z}_{\textrm{hori}}}\rangle = \langle\Pi^1_{\mathcal{Z}_{\textrm{vert}}}\rangle = 1$ and $\langle\Pi^1_{\mathcal{X}_{\textrm{hori}}}\rangle = 
	\langle\Pi^1_{\mathcal{X}_{\textrm{vert}}}\rangle = \frac{1}{3}$.
	\label{fig:gs}}
\end{figure*}

Here, we report the preparation of high-fidelity ground states of $\mathbb{Z}_3$ toric codes with periodic boundary conditions on up to $6 \times 4 = 24$ qutrits. We explicitly prepare two types of defects (Fig~\ref{fig:concept}): a parafermion defect (PF) and its conjugate ($\mathrm{PF}^*$)~\cite{fendley2012parafermionic, mong2014parafermionic, alicea2016topological,hutter2016quantum, fern20172} which, similar to the $\mathbb{Z}_2$-case, are related to dislocations of the lattice. Further, we prepare a charge conjugation (CC) defect, which has no analog in the qubit toric code and is related to  globally conjugating all charges and fluxes on-site. We verify their fusion rules by measuring their action on test anyons. Finally, we use these novel defects to produce a topological qutrit and initialise it by appropriately injecting long-range entanglement between two CC defect pairs. Experimentally, these advances are enabled by recent upgrades to Quantinuum's H2 ion-trap quantum computer allowing the use of 56 effectively all-to-all connected qubits, with two-qubit gate fidelities exceeding $99.8\%$~\cite{moses_race-track_2023, decross_computational_2024}. These gate fidelities allow us to encode qutrit degrees of freedom into the native qubits of the device (cf. Appendix \ref{app:circuit_decompositions}) and still achieve sufficiently high fidelities for the resulting one- and two-qutrit gates.

\section{Model \& Ground State Preparation}  

To initialize our experiments, we first prepare the ground state of the rotated $\ZZ_3$ toric code. The Hilbert space consists of qutrit degrees of freedom, namely $\ket{0}$, $\ket{1}$, and $\ket{2}$, on the vertices of a square lattice with periodic boundary conditions. Similar to the simpler case of the $\ZZ_2$ toric code, we define stabilizers $A = \mathcal{X}^\dagger \mathcal{X}  \mathcal{X}^\dagger \mathcal{X}$ and $B = \mathcal{Z} \mathcal{Z} \mathcal{Z}^\dagger \mathcal{Z}^\dagger$, which act non-trivially on alternating plaquettes of the lattice (Fig. \ref{fig:gs}a) ~\cite{kitaev2003fault,wen2003quantum}. $\mathcal{X}$ and $\mathcal{Z}$ correspond to the qutrit clock matrices:
\begin{align}
	\label{eq:XandZ}
	\mathcal{Z} \ket{i}= \omega^i \ket{i}
	\quad \mathrm{and} \quad
	\mathcal{X} \ket{i} = \ket{i + 1\ (\textrm{mod 3})}
\end{align}
where $\omega = e^{2 \pi i/3}$. While all $A$ and $B$ stabilizers commute due to the commutation relation $\mathcal{XZ} = \omega \mathcal{ZX}$, they are not Hermitian operators. To probe their expectation values, we consider the following projectors:
\begin{align}
	\label{eq:projectors}
	\Pi_A^{\alpha} &= \frac{1}{3}(\mathbf{1} + \alpha^2 A +  \alpha A^2) \quad \, \mathrm{and} \\ \nonumber
	\Pi_B^{\alpha} &= \frac{1}{3}(\mathbf{1} + \alpha^2 B +  \alpha B^2),
\end{align}
where $\alpha \in \{1,\omega, \bar \omega\}$.

The Hamiltonian for the $\mathbb{Z}_3$ toric code is defined as:
\begin{align}
	H = -\sum_{\substack{p\ \in\  \text{type-A} \\ \text{plaquettes}}} \Pi_{A_p}^1 - \sum_{ \substack{p \ \in\ \text{type-B} \\ \text{plaquettes}}} \Pi_{B_p}^1.
	\label{eq:tc_ham}
\end{align}
The ground state subspace of the Hamiltonian is the simultaneous +1-eigenspace of the $A_p$ and $B_p$ stabilizers. 
In analogy to the familiar case of $\ZZ_2$ toric code, stabilizer violations on plaquettes indicate the presence of anyons.  Specifically, a violation $\Pi_A^1 = 0$ (i.e., $A \neq +1$) on a type-A plaquette signals the presence of a charge anyon ($e$ or $\ebar$), while a violation $\Pi_B^1 = 0$ on a type-B plaquette indicates the presence of a flux anyon ($m$ or $\mbar$).
The anyon type can be determined by measuring $\Pi_A^{\omega}$ and $\Pi_A^{\bar \omega}$: $(\Pi_A^{\omega}, \Pi_A^{\bar \omega}) = (1,0)$ indicates the presence of an $e$ anyon, while $(\Pi_A^{\omega}, \Pi_A^{\bar \omega}) = (0,1)$ signifies an $\ebar$ anyon. Similarly, flux anyons $m$ and $\mbar$ can be distinguished by measuring $\Pi_B^{\omega}$ and $\Pi_B^{\bar \omega}$. 
On a torus, the ground state subspace of $H$ is spanned by nine degenerate states. These states can be distinguished by the logical string operators $\mathcal{Z}_{\textrm{hori}}$ and $\mathcal{Z}_{\textrm{vert}}$, which are products of qutrit $\mathcal{Z}$ and $\mathcal{Z^\dagger}$ operators that wrap around the torus in the horizontal and vertical directions, and take values $\{1, \omega, \bar \omega\}$.

To prepare the logical $\ket{00}_L$ ground state characterized by $ \mathcal{Z}_{\textrm{hori}}=+1=\mathcal{Z}_{\textrm{vert}}$, we use the protocol described in Ref.~\onlinecite{liu_methods_2022} and shown in Fig.~\ref{fig:gs}(b,c): The initial state of the $N$ qutrits in the quantum processor,  $\ket{0}^{\otimes N}$,  already fulfills $B = +1$ for all B-type plaquettes. We proceed by (i) choosing an A-type plaquette and a ``representative'' qutrit within it which is transformed into the state $\ket{+}:=\frac{1}{\sqrt{3}}(\ket{0}+\ket{1}+\ket{2})$, and (ii) applying a sequence of controlled-$\mathcal{X}$ (C$\mathcal{X}$) or C$\mathcal{X}^\dagger$ gates to the remaining qutrits within the plaquette, with the choice of gate (C$\mathcal{X}$ or C$\mathcal{X}^\dagger$) determined by whether the target qutrit is acted upon by $\mathcal{X}$ or $\mathcal{X}^\dagger$ in the stabilizer $A$~(Fig.~\ref{fig:gs}c). The action of the C$\mathcal{X}$ gate on two qutrits is $\textrm{C}\mathcal{X}\ket{i,j} = \ket{i,i+j\ (\text{mod 3})}$.  We repeat steps (i) and (ii) until all but one A-type plaquettes have been chosen, while carefully avoiding to designate a qutrit as ``representative'' that has previously been acted on by a C$\mathcal{X}$ gate (see Fig.~\ref{fig:gs}d for our chosen ordering). The final plaquette is implicitly prepared due to the symmetry constraint on the operators $\prod_p A_p = 1$.

At the end of the circuit, we measure all qutrits in both the $\mathcal{X}$ and $\mathcal{Z}$ bases to compute the expectation value of $\Pi^1_{\bullet}$ for every plaquette.  
A barrier is inserted before performing destructive qutrit measurements which ensures that the entire quantum state is  prepared before the measurements collapse the wavefunction into single qutrit eigenstates. As our qutrit encoding uses two qubits per qutrit, the remaining one-dimensional subspace can be used to detect errors during preparation that cause qutrits to leak outside of the qutrit subspace. These errors are heralded, and the corresponding shots are discarded, representing approximately 11\% of the total number of shots. The values presented in the main text are computed from the remaining, retained shots (cf. Appendix~\ref{app:gs_data} for the raw data for different system sizes).

To assess the quality of the prepared state, we show the expectation value of $\Pi_{\bullet}^1$ for each plaquette, as well as the logical operators in Fig.~\ref{fig:gs}(d,e). The logical mean values were computed by averaging across columns for horizontal operators and across rows for vertical operators. Measurement of the correlations between stabilisers of a given type allow us to bound the fidelity per site with the logical $\ket{00}_L$ state, detailed in Appendix~\ref{app:fidelity_bounds1}, as 
\begin{align}
    0.965(3) \leq \left(\bra{00}_L \rho \ket{00}_L\right)^{1/24} \leq 0.984(2)
\end{align}
and the lower bound further increases to 0.974(3) after accounting for readout errors, as discussed in Appendix~\ref{app:gs_data}. 

\begin{figure*}[t]\includegraphics[width=1.00\linewidth]{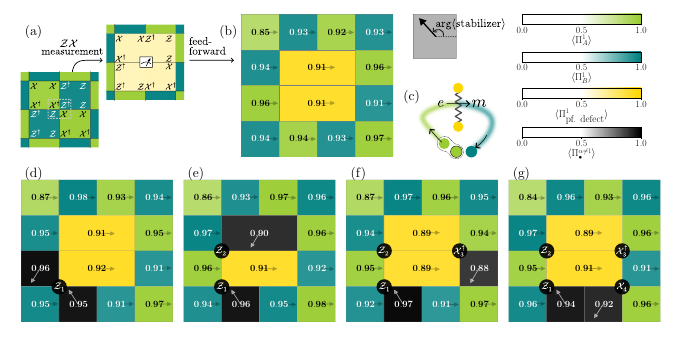}
\caption{\textbf{Creation of and braiding around parafermion defects.} 
    Any plaquette containing an anyon is colored black, with the value of $\max (\Pi_{\bullet}^{\omega}, \Pi_{\bullet}^{\bar \omega})$ displayed. An arrow within each plaquette indicates the direction specified by the arg $\langle\textrm{stabilizer}\rangle$, where the stabilizer could be $A_p$, $B_p$, or any of the defect stabilizers. The arrow's direction serves as a visual cue to distinguish anyons from their conjugates. (a,b) A pair of defects is inserted into the ground state by measuring the middle qutrit in the $\mathcal{X} \mathcal{Z}$-basis and performing feed-forward based on the measurement outcome. (c) A sketch illustrating the braiding experiment in steps (d-g). A pair of charges, $e$ and $\ebar$, is created. Charge $\ebar$ remains fixed, while $e$ is dragged through the defect pair  and emerges as $m$ on the other side of the defect pair. The maximum estimation error is 0.022.}    \label{fig:em_defect}
\end{figure*}

We observe that the expectation value of $\Pi_{A_p}^1$ for the implicitly prepared plaquette (i.e., the bottom right plaquette in Fig.~\ref{fig:gs}d) is generally slightly lower than the expectation values of the remaining type-A plaquettes, which are explicitly prepared. This observation can be attributed to the fact that, if a charge pair ($e-\ebar$) is created due to a $\mathcal{Z}$ or $\mathcal{Z}^\dagger$ error during state preparation, the sequential preparation of plaquettes will drag one of these spurious anyons all the way to final plaquette. This is in contrast to non-unitary preparation schemes where no significant translation symmetry breaking has been observed even when subjected to noise \cite{iqbal_topological_2024, foss2023experimental}.

\section{Parafermion Defects}
Equipped with a high-fidelity ground state, we turn to the study of topological defects. One type of defect supported by the $\mathbb{Z}_3$ toric code generalises the well known $em$-defect of its $\mathbb{Z}_2$ counterpart. That defect acts on the flux and charge anyons by exchanging their identities and behaves like a Majorana~\cite{kitaev2006anyons,bombin2010topological}, which has been used experimentally to implement logical Clifford gates on a $\mathbb{Z}_2$ toric code background~\cite{google2023non, xu2023digital}. In contrast, for the  $\mathbb{Z}_3$ toric code, a generalisation of such a defect has been predicted to have parafermion fusion rules~\cite{you2012projective}
\begin{equation}
    \mathrm{PF} \times \mathrm{PF} = \vac + \ebar m + e \mbar.
\label{eq_pf_fusion}
\end{equation}
As we will see, two distinct species of parafermions, labeled PF and PF$^*$ can arise in the $\mathbb{Z}_3$ case.

\begin{figure*}[t]
	\includegraphics[width=0.95\linewidth]{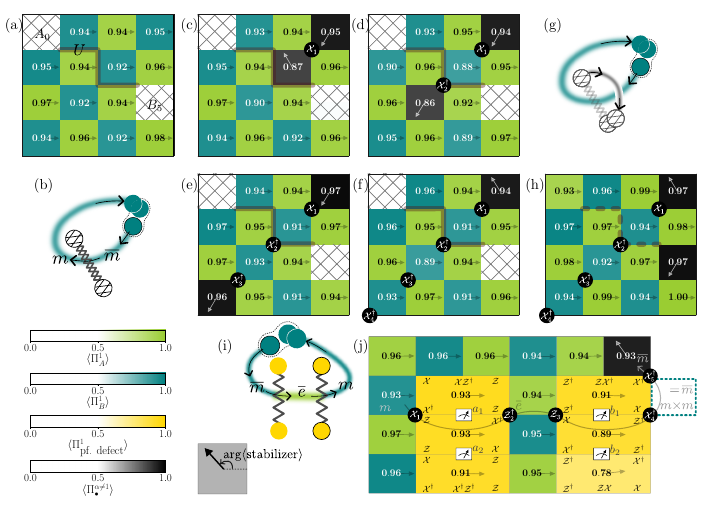}
	\caption{\textbf{Creation and braiding of CC defects and their relation to parafermions.} (a) Ground state of the $\mathbb{Z}_3$ toric code with a CC defect pair. The endpoints of the thick line, representing the CC construction circuit $U$, correspond to high-weight stabilizers $A_0$ and $B_5$ (defined in Fig.~\ref{fig:cc_construction}c). These stabilizers are marked by a hatching pattern with a `$\times$' symbol, and their values are omitted for clarity as they label the internal state of the defects which is not locally accessible.  (b) A sketch of the braiding experiment in (c-f). A flux pair $m - \mbar$ is created.  $\mbar$ is transmuted into $m$ by commuting it through the CC defect line, and it is then fused with the fixed $m$ anyon at the top right corner through a sequence of four steps (c-f), by applying $\mathcal{X}_1$, $\mathcal{X}_2^\dagger$, $\mathcal{X}_3^\dagger$, and  $\mathcal{X}_4^\dagger$. (g) Outline of the braiding experiment in (h) where the CC defect pair is fused. This is achieved by applying the same circuit $U$ used in (a) (see Appendix \ref{app:coherently_moving} for derivation). The altered state of the CC defect pair is revealed as a flux $m$ at one endpoint. (i) A sketch of the braiding experiment in (j). We prepare the ground state and create two parafermion defect pairs. The $m$ flux from the pair $m-\mbar$ created at the top left corner remains fixed, while its partner $\mbar$ anyon is commuted through two parafermion defect pairs. The resulting $m$ is then fused with the pinned $m$ to give a single $\mbar$ anyon.		 
	\label{fig:cc-defect}}
\end{figure*}

To prepare a pair of PF defects we first initialise the ground state and then measure one of the qutrits in the basis in which either the operator $\mathcal{XZ}$ or $\mathcal{XZ^\dagger}$ (corresponding to the distinct parafermion species) is diagonal.
To initialize the defect pair in a well-defined state, we feed-forward on the measurement outcome which ensures that the corresponding defect stabilizers have definite values (Fig. \ref{fig:em_defect}a). Having created the ground state with an initialized em-defect pair (cf. Fig. \ref{fig:em_defect}b), we create a charge-anticharge ($e-\ebar$) pair and subsequently braid $e$ around one of the defects using the qutrit clock matrices (Fig. \ref{fig:em_defect}(d-g)). Upon passing through the line connecting the defects, a change in stabilizer expectation values indicates that a permutation
\begin{equation}
    {e} \rightarrow {m}
\label{eq_pf_action}
\end{equation} has occurred. Finally, at the end of the experiment, we are left with a single dyon $\ebar m$, i.e. the fusion outcome in~\eqref{eq_pf_fusion} has been toggled from the identity to the dyon channel. This demonstrates the parafermion behavior predicted for the $\ZZ_3$ toric code.

\section{Charge Conjugation Defects and Relation to Parafermions}
The richer anyon content of the $\ZZ_3$ toric code allows for another type of topological defect, which has no analog in the $\ZZ_2$ case: As an anyon traverses a line connecting a pair of such ``charge conjugation'' (CC) defects, it is turned into its antiparticle, i.e., $e \leftrightarrow \ebar$ and $m \leftrightarrow \mbar$. Experimentally demonstrating the action of this novel type of defect is what we turn to next.

To this end, we apply a circuit $U$ (cf.~Fig.~\ref{fig:cc_construction}b) to the ground state of the $\mathbb{Z}_3$ toric code. A key ingredient of the CC defect pair unitary construction is the one-qutrit ``charge conjugation'' gate, which acts as $\mathcal{C} \ket{i} = \ket{-i \ (\text{mod 3}) }$.
Intuitively, the circuit construction  ``unzips'' the $\mathbb{Z}_3$ toric code to a trivial paramagnet, applies $\mathcal{C}$ to this trivial state, and then returns to the $\mathbb{Z}_3$ toric code---see Appendix~\ref{app:z3_defects} for a more in-depth derivation \cite{lyons-2024}. 
We then create a pair of $m - \mbar$ anyons and move $\mbar$ through the line connecting the CC defect pair. This operation transforms
\begin{equation}
	\mbar \rightarrow m,
	\label{eq_cc_action}
\end{equation}
as evidenced by the change in the direction given by arg $\langle \text{stabilizer} \rangle$, which is visually represented by an arrow within each plaquette in Fig.~\ref{fig:cc-defect}. The arrow direction changes approximately from $120^\circ$ to $240^\circ$ in the excited plaquettes on two sides of the CC defect line (Fig.~\ref{fig:cc-defect}(c,d)). The transformed anyon ($m$) is then transported around the torus and fused with its partner (another $m$), resulting in a single $\mbar$ particle (Fig.~\ref{fig:cc-defect}(e,f)). 

Crucially, as the $\mbar$ particle traverses the defect line, it alters the internal state of the CC defect pair which is invisible for local observables. This altered state manifests itself as an $m$ anyon (cf. Fig.~\ref{fig:cc-defect}h) upon coherently moving and fusing back the CC defect pair by applying the same unitary $U$ used in Fig.~\ref{fig:cc_construction}b (cf. Appendix~\ref{app:coherently_moving} for why applying $U$ again achieves coherent movement of the end of a CC defect), demonstrating an instance of the fusion rules
\begin{align}
    \mathrm{CC} \times \mathrm{CC} = (\vac +  \ebar m + e\mbar) (\vac + em + \ebar\mbar).
    \label{eq_cc_cc_fusion}
\end{align}
We will later exploit the ability of the CC defect pair to store information in its fusion channel, which allows for the distribution of entanglement in a non-local way.

\begin{figure*}[t]
	\centering    \includegraphics[width=0.99\linewidth]{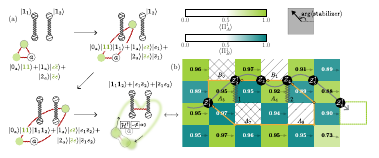}
	\caption{\textbf{Entanglement transfer from anyons to CC defects for initialising topological qutrits.} (a) A sketch of the different steps, with intermediate states, involved in moving a charge anyon around defects. The braiding followed by measuring an ancilla transfers a Bell state of charge anyons into an entangled logical state of CC defects. (b) Results for the final step as depicted in (a). We create $\ZZ_3$  ground state with two pairs of defects, labeled 1 and 2 (cf. Figure~\ref{fig:6x4_entangle_defects_construction}). Defect pair 1, marked with the $\times$-hatch pattern, extends between points $B_0$ and $A_7$. Defect pair 2, indicated by the $/$-hatch pattern, has endpoints $B_1$ and $A_8$.  The orange loop represents a braid that stabilizes the prepared topological qutrit state; a black border on the solid circle indicates the application of $\mathcal{X}^\dagger$, while its absence indicates $\mathcal{X}$. The expectation values of the projectors $\Pi_{B_0}^1$ and $\Pi_{B_1}^1$ for the non-local stabilizers are $0.931(15)$ and $0.927(15)$, respectively. 	We measure the expectation values of $\Pi_{A_7}^1$, $\Pi_{A_8}^1$, and $\Pi_{A_7A_8}^1$ for different ancilla outcomes ($|0_a\rangle$, $|1_a\rangle$, and $|2_a\rangle$). For $\Pi_{A_7}^1$, the measured values are 0.44(5), 0.40(5), and  0.39(5) respectively. Similarly, for $\Pi_{A_8}^1$, the values are 0.44(5), 0.42(5), and 0.36(5). Finally, $\Pi^1_{A_7A_8}$ yields values of 0.81(4), 0.78(4), and 0.74(4) for the respective ancilla states. This is a manifestation of the fact that, although the outcomes for each individual defect pair are random, they are jointly in an entangled state.\label{fig:entangle_defects}}
\end{figure*}

Having demonstrated the action of parafermion~\eqref{eq_pf_action} and CC defects~\eqref{eq_cc_action} on the anyons, we are now in a position to demonstrate their mutual fusion rule, namely,
\begin{equation}
    \mathrm{PF} \times \mathrm{PF}^* = \mathrm{CC},
\label{eq_pf_pf_fusion}
\end{equation}
i.e., the combined action of a parafermion defect and its conjugate is equivalent to that of a charge conjugation. Note that~\eqref{eq_pf_pf_fusion} is compatible with multiplying~\eqref{eq_pf_fusion} with its conjugate and comparing to~\eqref{eq_cc_cc_fusion}. To illustrate this, we generate two pairs of parafermion defects (cf. Fig.~\ref{fig:cc-defect}j). 
For the first pair, we measure the qutrit $a_1$ in the $\mathcal{XZ}$ basis and the qutrit $a_2$ in the $\mathcal{XZ}^\dagger$ basis, resulting in a pair of PF defects. Similarly, for the second pair, we measure qutrit $b_1$ in the $\mathcal{XZ}$ basis and qutrit $b_2$ in the $\mathcal{XZ}^\dagger$ basis, producing a pair of PF* defects. Qutrit measurements are followed by a real-time feedforward operation to deterministically initialize the parafermion defects. Next, we act with $\mathcal{X}_1$, which creates an $m$ anyon in a type-A plaquette and simultaneously excites the inner defect plaquette of the left parafermion defect pair (Fig.~\ref{fig:cc-defect}j). We then move one of the anyons to the right, wrapping around the torus. The left parafermion defect pair transmutes $\mbar$ to $\ebar$, while the right parafermion defect pair changes $\ebar$ to $m$. This $m$ anyon is then fused with its partner, the original $m$ anyon, leaving us with a single $\mbar$ anyon. This is the exact same outcome we obtained for a similar anyon trajectory and a single pair of CC defects. In addition, Supplementary Fig.~\ref{fig:pf-conjugate} partially demonstrates the conjugation of a parafermion defect line with a CC defect line, effectively resulting in a conjugate parafermion defect line. 

\section{A Topological Qutrit}
\label{sec:topo_qutrit}
Finally, having established control over the anyons and defects of the $\mathbb{Z}_3$ toric code, we use these ingredients to produce a topological qutrit. Ref.~\onlinecite{levaillant_universal_2015} gives the standard definition of a topological qutrit as a collection of four non-Abelian quasiparticles whose overall fusion outcome is neutral but individual pairs have three possible fusion outcomes.

The non-Abelian objects we use to create such a qutrit are given by two pairs of charge conjugation defects, created from the toric code vacuum. In principle, a pair of defects can have nine different fusion outcomes, according to Eq.~\eqref{eq_cc_cc_fusion}, representing two qutrits worth of information. For simplicity, here we will focus on the single-qutrit subspace spanned by
\begin{equation}
\begin{aligned}
\ket{\phi_{1}} &= \Big[ \ket{\vac_1 \vac_2} + \ket{e_1 \ebar_2} +   \ket{ \ebar_1 e_2} \Big]/\sqrt{3} \\ \ket{\phi_{\omega}} &= \Big[\ket{\vac_1 \vac_2} + \omega \ket{e_1 \ebar_2} + \bar \omega \ket{ \ebar_1 e_2} \Big]/\sqrt{3} \\
\ket{\phi_{\bar \omega}} &= \Big[ \ket{\vac_1 \vac_2} + \bar \omega \ket{e_1 \ebar_2} + \omega \ket{ \ebar_1 e_2}  \Big]/\sqrt{3}
\label{eq_topological_qutrit_states}
\end{aligned}
\end{equation}
where $\ket{\alpha_1 \beta_2}$ denotes the state in which the left (right) defect pair fuses to $\alpha$ ($\beta$). That is, we restrict to the magnetically neutral sector in which no $m$ or $\bar{m}$ anyons appear in any of the intermediate fusion outcomes.

The logical $Z_L$ and $X_L$ operators on this topological qutrit can be realised on the physical level as follows: Creating a charge-anti-charge pair and moving the charge through both defect lines (e.g., using the path given by the gray line in Fig.~\ref{fig:entangle_defects}b)  implements a logical $Z_L = \ket{\vac_1 \vac_2} \bra{e_1 \ebar_2} + \ket{e_1 \ebar_2} \bra{\ebar_1 e_2} + \ket{\ebar_1 e_2} \bra{\vac_1 \vac_2}$. Similarly, braiding a flux around \emph{one} of the defect pairs picks up a phase depending on the internal state of that defect and thus this operators realises a logical $X_L = \ket{\phi_{1}} \bra{\phi_{\omega}} + \ket{\phi_{\omega}} \bra{\phi_{\bar \omega}} + \ket{\phi_{\bar \omega}} \bra{\phi_{1}}$.

However, $X_L$ or $Z_L$ cannot initialise the topological qutrit states~\eqref{eq_topological_qutrit_states} starting from $\ket{\vac_1 \vac_2}$. To do this, we demonstrate the application of a logical Fourier transform $\mathcal{H}$ (cf. Appendix~\ref{app:circuit_decompositions}): A superposition of states with physical anyons pairs $e\ebar$, and $\ebar e$ injected at a fixed position and the vacuum is produced using a control-$\mathcal{Z}$ operation conditioned on an ancilla qutrit `$a$' initialized in the state $(\ket{0}_a + \ket{1}_a + \ket{2}_a)/\sqrt{3}$. One of the anyons is then coherently moved to braid around one half of each defect pair before being annihilated with its partner. The intermediate states of the ancilla, charge anyons, and defect pairs are depicted in Fig.~\ref{fig:entangle_defects}a. After applying an additional $\mathcal{H}^\dagger$ gate (cf. Appendix~\ref{app:circuit_decompositions}) on the ancilla, the resulting state of defect pairs and ancilla is proportional to:
\begin{equation}
\begin{aligned}
	 \ket{0}_a \ket{\phi_{1}} + \ket{1}_a \ket{\phi_{\omega}} + \ket{2}_a &\ket{\phi_{\bar \omega}}.
\label{eq_bell_pairs}
\end{aligned}
\end{equation}
After measuring the ancilla and recording the measurement outcome, we have prepared the logical state $\ket{\phi_{\omega^j}}$ where $j$ is the measurement outcome of the ancilla. A logical $X_L$ operation can be used to complete the state preparation protocol if deterministic state preparation is desired.

Crucially, the entanglement between the system and the ancilla has been transferred from a local to a non-local information carrier and is now robust: As long as the distance between both the endpoints of each of the defects as well as between the defect lines themselves is sufficiently large, any anyons created by a local noise process can maximally encircle a single endpoint. However, any process in which there is an odd number of anyons crossings the defect lines will result in an odd number of charge or flux anyons. By fusing the spurious anyon back into the closest defect, we can return to the original logical state.

To certify the non-local entanglement of the defect pairs, we focus on the shots where the ancilla has been measured in the $\ket{0}$ state and make use of the fact that 
$\ket{\phi_1}$ is uniquely specified by being a +1 eigenstate of two commuting anyon braids: The first braids a flux around \textit{both} defect pairs (denoted by the orange line in Fig.~\ref{fig:entangle_defects}b) and is microscopically implemented by a string of physical $\mathcal{X}$ and $\mathcal{X}^\dagger$.
The second anyon braid is simply the logical $Z_L$-operator defined above. Measuring the expectation values of these operators leads to  fidelity bounds

\begin{equation}
    0.72(5) \leq \mathrm{Tr}[\bra{\phi_1} \rho \ket{\phi_1}] \leq 0.80(4),
\end{equation}
and we report the results for the different states $\ket{\phi_{\omega^j}}$ as well as SPAM-corrected values in 
Appendix~\ref{app:fidelity_bounds}. All states are prepared with fidelities that far exceed those that can be reached with a classical mixture, which certifies that the two fusion channels of the defect pairs have been successfully entangled. For comparative analysis, Appendix~\ref{app:cc-entanglement-6x2} presents results for a $6\times2$ lattice geometry, providing a more detailed description of the entire procedure. 
We note that other logical states, like $\ket{\phi_0} + \ket{\phi_\omega} + \ket{\phi_{\bar \omega}} \propto \ket{\vac_1 \vac_2}$ are easier to achieve than what we have presented since they do not require the ancilla-assisted logical Fourier transform gate we have executed.

\section{Conclusion}
We have created high-fidelity ground states of the $\ZZ_3$ toric code on tori of sizes up to $6\times 4 = 24$ qutrits using Quantinuum's H2 trapped ion quantum computer. We have created parafermion and charge conjugation defect pairs on top of this vacuum and verified their action on the anyons of the model; some of these operations were facilitated by using adaptive quantum circuits. Finally, we have initialised a topological qutrit from two charge conjugation defect pairs. An appealing direction for future work is to explore syndrome measurements and repeat-until-success protocols, which have thus far led to a break-even for state preparation and measurement errors for simpler code states \cite{egan2021faulttolerantoperationquantumerrorcorrection, acharya_quantum_2024,putterman2024hardwareefficientquantumerrorcorrection,Sivak_2023, bluvstein_logical_2024}. 

Our findings show that digital quantum processors have advanced to the point where they can throw off the shackles of the underlying qubit architecture and explore the much larger class of topologically ordered systems that is naturally formulated with local Hilbert space dimensions greater than two. Tantalisingly, that class includes models with universal quantum computational power, some of which are closely related to the model presented here~\cite{mochon_anyon_2004}.

\section*{Data availability}
The numerical data that support the findings of this study are available from Zenodo repository 10.5281/zenodo.14007593~\cite{supporting_2024}.

\section*{Code availability}
The code used for numerical simulations is available from from Zenodo repository 10.5281/zenodo.14007593~\cite{supporting_2024}.

\section*{Author Contributions}
J.D., C.F., T.M.G., D.G., N.H., C.A.H., J.J., B.N., Y.M., M.M., S.A.M. and P.S. ran the experiment and took the data. M.I., A.L., C.F.B.L., N.T., A.V., R.V. and H.D. conceived the experiments, translated the ideas to quantum circuits, did the data analysis and drafted the manuscript.

\section*{Acknowledgements}
We thank the broader team at Quantinuum for comments. N.T. is supported by the Walter Burke Institute for Theoretical
Physics at Caltech. A.V. and R.V. are supported by the Simons Collaboration on Ultra-Quantum Matter, which is a grant
from the Simons Foundation (618615, A.V.). A.L. and C.F.B.L. acknowledge support from the National Science Foundation Graduate Research Fellowship Program (NSF GRFP).
This work is in part supported
by the DARPA MeasQuIT program.
The experimental data in this work was produced by the Quantinuum
H2 trapped ion quantum computer, powered by Honeywell, in 2024.

\bibliography{references, references_henrik}

\onecolumngrid
\newpage

\appendix

\section{Details about the Circuit Decompositions}
\label{app:circuit_decompositions}

This section details the construction of primitive gates for manipulating $\ZZ_3$ qutrits, utilizing the native gate set of Quantinuum's H2-series devices \cite{h2-data-sheet, noauthor_quantinuum_2024}. A key consideration in circuit design is the number of two-qubit ZZPhase gates (the native entangling gate) required in each decomposition, as this directly influences the implementation cost. We encode each qutrit degree of freedom into two qubits according to the mapping:
\begin{align}
	\label{eq:encoding}
	\ket{0}_{\textrm{qutrit}}:=\ket{00}, \quad
	\ket{1}_{\textrm{qutrit}}:=\ket{10}, \quad
	\ket{2}_{\textrm{qutrit}}:=\ket{11}, \quad
	\ket{\textrm{nc}} := |01\rangle
\end{align}
The non-computational state $\ket{\textrm{nc}}$ is used to herald errors that cause the two encoding qubits to fall outside the qutrit space. Given that we have following single- and two-qubit gates natively available on the device, 
\begin{equation}
	\label{eq:1qgates}
	U_{1q}(\theta,\phi) = e^{-i(\cos \pi \phi X + \sin \pi\phi Y)\frac{\theta}{2}} := \cbox{4.0}{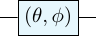}  
	, \quad 
	R_Z(\theta) = e^{-\frac{1}{2}i\pi\theta Z},
\end{equation}
\begin{equation}
	\label{eq:ZZphase}
	\textrm{ZZPhase}(\theta) = e^{-\frac{1}{2}i\pi\theta (Z\otimes Z)}.
\end{equation}
The action of the qutrit $\mathcal{Z}$, defined in \eqref{eq:XandZ}, on the encoded space can be implemented efficiently using two single-qubit gates.
\begin{align}
	\label{eq:Z}
	\mathcal{Z} &:= 
	\cbox{5}{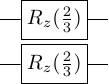} 
	= \bar \omega\ \textrm{diag}(1,\ \omega,\ \omega,\ \bar \omega)
\end{align}
Importantly the action of $\mathcal{Z}$ on $|\textrm{nc}\rangle$ is trivial up to the phase.
In contrast, implementing $\mathcal{X}$ requires two CNOT gates, or a single ZZPhase gate.
\begin{align}
	\label{eq:qutrit-Z}
	\mathcal{X} &:= \cbox{5}{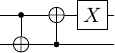} 
	= \cbox{20}{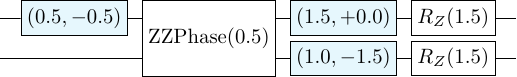} 
	=     \begin{pmatrix}
		0 & 0 & 0 & 1\\
		1 & 0 & 0 & 0\\
		0 & 0 & 1 & 0\\
		0 & 1 & 0 & 0\\
	\end{pmatrix}
\end{align}
Similarly, the charge conjugation gate $\mathcal{C}$ can also be implemented using only a single ZZPhase gate. 
\begin{align}
	\label{eq:qutrit-C}
	\mathcal{C} &:=
	\cbox{5}{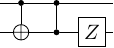} 
	= \cbox{20}{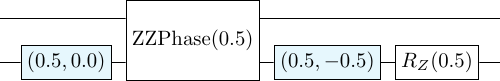} 
	=  \begin{pmatrix}
		1 & 0 & 0 & 0\\
		0 & 0 & 0 & 1\\
		0 & 0 &-1 & 0\\
		0 & 1 & 0 & 0
	\end{pmatrix}
\end{align}
Another primitive gate for our purposes is the qutrit Fourier transform, denoted by $\mathcal{H}$, which acts on computational states as follows: 
\begin{align}
	\label{eq:qutrit-H}
	\mathcal{H} \ket{i}_\text{qutrit} = \frac{1}{\sqrt{3}} (\ket{0}_\text{qutrit} + \omega^i \ket{1}_\text{qutrit} + \omega^{2i} \ket{2}_\text{qutrit}),
\end{align}	
while leaving $\ket{\textrm{nc}}$ unchanged. The native implementation of $\mathcal{H}$ requires three ZZPhase gates. 

To construct C$\mathcal{X}$, we begin by implementing C$\mathcal{Z}$, which utilizes four Controlled-$R_Z$ gates. Applying the C$\mathcal{Z}$ gate to a target qutrit transformed by $\mathcal{H}$, results in the required C$\mathcal{X}$ gate. 
\begin{align}
	\label{eq:qutrit-CX}
	\textrm{C}\mathcal{X} &:=
	\cbox{26}{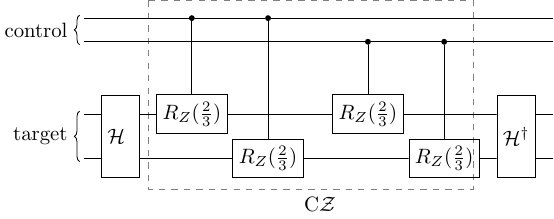} 
\end{align}
We observe that the action of $\mathcal{H}$ on the $\ket{0}_{\textrm{qutrit}}=\ket{00}$ results in the superposition $\frac{1}{\sqrt3}(\ket{00} + \ket{10} + \ket{11})$. We can consider $\mathcal{H}\ket{00}$ as a state preparation procedure that can be implemented using only a single ZZPhase gate \cite{iten2016quantum}. 
\begin{align}
	\label{eq:state_prep}
	\mathcal{H}\ket{00} &=
	\cbox{40}{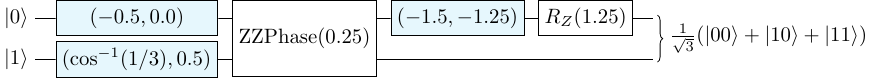} 
\end{align}

Preparing the $\mathbb{Z}_3$ ground state on a $6\times4$ lattice using Quantinuum's H2-series native gateset requires 251 two-qubit gates for final destructive measurements in the
$\mathcal{Z}$-basis and 189 two-qubit gates for measurements in the $\mathcal{X}$-basis.  The circuit depth is 113 for $\mathcal{Z}$-basis measurements and 106 for $\mathcal{X}$-basis measurements.

\clearpage

\section{Fidelity Bounds}
\label{app:fidelity_bounds}
This appendix provides detailed derivations of fidelity bounds for both the $\ZZ_3$ ground state and topological qutrit by entangling CC defect pairs. 
In Sec.~\ref{app:fidelity_bounds1}, we focus on establishing the bounds for the $\ZZ_3$ ground state, and in Sec.~\ref{app:fidelity_bounds2}, we utilize similar reasoning to derive the corresponding bounds for topological qutrit states.

\subsection{$\ZZ_3$ Ground State}
\label{app:fidelity_bounds1}
The projector onto the logical $\ket{00}$ ground state can be written as a product of projectors
\begin{equation}
\ket{00}\bra{00} = PQ
\end{equation}
with
\begin{equation}
\begin{aligned}
P &= \prod_{\substack{x\ \in\  \text{type-A} \\ \text{plaquettes}}} \Pi_{A_x}^1 \\ 
Q &= \left(\prod_{ \substack{z \ \in\ \text{type-B} \\ \text{plaquettes}}}  \Pi_{B_z}^1 \right) \Pi^1_{\mathcal{Z}-\mathrm{hori}} \Pi^1_{\mathcal{Z}-\mathrm{vert}}.
\end{aligned}
\end{equation}
Here, $x$ and $z$ run over all plaquettes of the given type except one and the geometry of the logical operators $\Pi^1_{\mathcal{Z}-\mathrm{hori}}$ and $\Pi^1_{\mathcal{Z}-\mathrm{vert}}$ can be chosen arbitraily --- each choice leads to a valid bound.

Since $[P,Q] = 0$, we can label a complete set of states by their eigenvalues with respect to $P$ and $Q$ (denoted $p$ and $q$ which are each 0 or 1) such that we can decompose the density matrix of the prepared state as a convex combination
\begin{equation}
\begin{aligned}
    \rho &= \sum c_{pq} \rho_{pq} \\ 
    &= c_{00} \rho_{00} +
    c_{01} \rho_{01} +
    c_{10} \rho_{10} +
    c_{11} \rho_{11}.
\end{aligned}
\end{equation}
Here, $c_{11}$ is the target state fidelity that we are seeking to bound. Furthermore, we have 
\begin{equation}
\begin{aligned}
    \mathrm{Tr}[\rho P] &= c_{10} + c_{11} \\
    \mathrm{Tr}[\rho Q] &= c_{01} + c_{11} \\ 
    1 &= c_{00} + c_{01} + c_{10} + c_{11} 
\end{aligned}
\end{equation}
from which we conclude our lower bound
\begin{equation}
\begin{aligned}
    c_{11} &= \mathrm{Tr}[\rho P] + \mathrm{Tr}[\rho Q] - 1 + c_{00} \\
    &\geq \mathrm{Tr}[\rho P] + \mathrm{Tr}[\rho Q] - 1.
\end{aligned}
\end{equation}
On the other hand, we have as an upper bound
\begin{equation}
\begin{aligned}
    c_{11} &\leq \min\{ \mathrm{Tr}[\rho P], \mathrm{Tr}[\rho Q]\}.
\end{aligned}
\end{equation}
For the $6 \times 4 = 24$ qutrit system, we have measured
\begin{equation}
\begin{aligned}
        \mathrm{Tr}[\rho P] &= 0.75(2) \\
        \mathrm{Tr}[\rho Q] &= 0.68(3)
\end{aligned}
\end{equation}
for the global projectors. From this, the bound for the global fidelity and fidelity per qutrit
\begin{equation}
\begin{aligned}
    \bra{00} \rho \ket{00} &\in [0.42(4), 0.68(3)] \\
    \bra{00} \rho \ket{00}^{1/24} &\in [0.965(3),  0.984(2)]
\end{aligned}
\end{equation}
follows. Employing SPAM error correction, as discussed in Appendix~\ref{app:gs_data}, we obtain improved fidelity bounds:
\begin{equation}
\begin{aligned}
	\bra{00} \rho \ket{00}_{\textrm{SPAM error mitigated}} &\in [0.53(3), 0.73(3)] \\
	\bra{00} \rho \ket{00}^{1/24}_{\textrm{SPAM error mitigated}} &\in [0.974(3), 0.987(1)].
\end{aligned}
\end{equation}
\subsection{Topological Qutrit States}
\label{app:fidelity_bounds2}
The exact same argument is used to bound the fidelity of topological qutrits encoded in the entangled states of two pairs of charge conjugation defects. For example, in the case where the post-measurement state of the ancilla is $|0\rangle$, the state $|{\vac \vac}\rangle + |e\ebar\rangle + |\ebar e\rangle$ is prepared. In this case, $P$ is the projector onto the +1 eigenspace of the operator denoted by $\Xent \Xent^\dagger$, where $\Xent$ changes the internal states according to: 
\begin{equation}
	\begin{aligned}
		\label{eq:x_ent}
		\Xent = \ket{\vac }\bra{e} + \ket{e}\bra{\ebar} +\ket{\ebar}\bra{\vac}.
	\end{aligned}
\end{equation}
Note that $\Xent \Xent^\dagger=Z_L$, as defined in Sec.~\ref{sec:topo_qutrit}, and the gray loop in Fig.~\ref{fig:entangle_defects}b depicts the concrete implementation of this operator in terms of $\mathcal{Z}$ and $\mathcal{Z}^\dagger$.  Similarly, $Q$ is the projector onto the +1 eigenspace of the $\Zent \Zent$ operator (as depicted by the orange loop in Fig.~\ref{fig:entangle_defects}b), and $\Zent$ acts like:
\begin{equation}
	\begin{aligned}\label{eq:z_ent}
		\Zent = \ket{\vac}\bra{\vac} + \omega \ket{e}\bra{e} +\bar \omega\ket{\ebar}\bra{\ebar}.
	\end{aligned}
\end{equation}

Moreover, $\Xent \Xent^\dagger$ distinguishes the three topological qutrit states with eigenvalues of +1, $\omega$, and $\bar \omega$. The expectation values of the gray and orange string loops are measured to obtain the values for the topological qutrit stabilizers $\Xent \Xent^\dagger$ and $\Zent \Zent$. We measure the following values:

\begin{center}
	\setlength{\tabcolsep}{3pt} 
	\renewcommand{\arraystretch}{1.7} 
	\begin{tabular}{c c c c} 
		\hline 
		Ancilla outcome & State of defect pairs & (${\Pi^1_{\Xent \Xent^\dagger}}$, ${\Pi^{\omega}_{\Xent \Xent^\dagger}}$, ${\Pi^{\bar \omega}_{\Xent \Xent^\dagger }}$) & 
		(${\Pi^1_{\Zent \Zent}}$, ${\Pi^{\omega}_{\Zent \Zent}}$, ${\Pi^{\bar \omega}_{\Zent \Zent}}$) \\[.5ex] 
		\hline 
		0 & $|\phi_1\rangle=\ket{\vac\vac} + |e \ebar\rangle + |\ebar_1e_2\rangle$ 
		& (0.92(3), 0.07(3), 0.01(1))  & (0.80(4), 0.09(3), 0.11(3)) \\ 
				
		1 & $|\phi_{\omega}\rangle=\ket{\vac\vac} + \omega |e \ebar\rangle + \bar \omega |\ebar_1e_2\rangle$ 
		& (0.009(9), 0.94(2), 0.05(2))  & (0.75(4), 0.13(3), 0.12(3)) \\  

		2 & $|\phi_{\bar \omega}\rangle=\ket{\vac\vac} + \bar \omega |e\ebar\rangle + \omega |\ebar_1e_2\rangle$ 
		& (0.06(2), 0.02(1), 0.92(3))  & (0.68(5), 0.11(3), 0.20(4)) \\ 
		[0.5ex] 
		\hline
	\end{tabular}
\end{center}

Repeating the argument above for each of the three post-measurement states and of the ancilla corresponding topological qutrit states $\ket{\phi_1}, \ket{\phi_\omega}$ and $\ket{\phi_{\bar \omega}}$, we obtain the bounds:
\begin{equation}
	\begin{aligned}
		\mathrm{Tr}[\bra{\phi_{1}} \rho \ket{\phi_{1}}] &\in [0.72(5), 0.80(4)]&\quad & \textrm{if ancilla in state} \ket{0} \\
		\mathrm{Tr}[\bra{\phi_\omega} \rho \ket{\phi_\omega}] &\in [0.70(5), 0.75(4)]  &\quad & \textrm{if ancilla in state} \ket{1} \\
		\mathrm{Tr}[\bra{\phi_{\bar \omega}} \rho \ket{\phi_{\bar \omega}}] &\in [0.60(6), 0.68(5)] &\quad & \textrm{if ancilla in state} \ket{2}.
	\end{aligned}
\end{equation}
Applying SPAM error mitigation, we again obtain improved fidelities:
\begin{equation}
	\begin{aligned}
		\mathrm{Tr}[\bra{\phi_1} \rho \ket{\phi_1}]_{\textrm{SPAM error mitigated}} &\in [0.76(4), 0.83(4)] &\quad & \textrm{if ancilla in state} \ket{0} \\
		\mathrm{Tr}[\bra{\phi_\omega} \rho \ket{\phi_\omega}]_{\textrm{SPAM error mitigated}} &\in  [0.73(5), 0.77(4)] &\quad & \textrm{if ancilla in state} \ket{1} \\
		\mathrm{Tr}[\bra{\phi_{\bar \omega}} \rho \ket{\phi_{\bar \omega}}]_{\textrm{SPAM error mitigated}} &\in [0.63(5), 0.70(5)] &\quad & \textrm{if ancilla in state} \ket{2}.
	\end{aligned}
\end{equation}
Each of these fidelities is significantly larger than the maximum fidelity achievable with a classical mixture.

\clearpage
\section{Ground State Preparation Data Analysis}
\label{app:gs_data}
\begin{figure*}[!ht]
\centering    \includegraphics[width=1.00\linewidth]{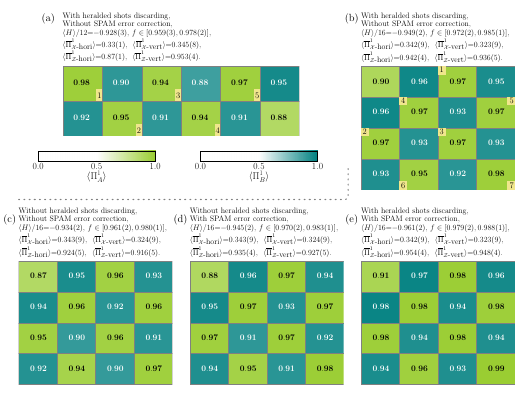}
\caption{\textbf{Ground state energy densities} Expectation values for stabilizers $\Pi_A^1$ and $\Pi_B^1$ on a torus. (a,b) We present preparation data for $6\times2$ and $4\times4$ lattices. For each type-A plaquette, a small square indicates the location of the control qutrit used in its preparation. The number within the square denotes the order in which the plaquettes were prepared. We discard heralded shots where qutrits are measured outside the qutrit space, while no other error mitigation technique is employed. (c,d,e) We compare the impact of heralded shot discarding and SPAM error mitigation on the expectation values of energy densities for the same 4x4 lattice as in (b)
\label{fig:gs_smaller_sizes}}
\end{figure*}

This appendix presents additional findings regarding the preparation of the $\mathbb{Z}_3$ ground state. 
We show the expectation values for the prepared ground state on a $6\times2$ and $4\times4$ lattice (cf. Fig.~\ref{fig:gs_smaller_sizes}(a,b)). 
These values were computed by discarding heralded shots, similar to the data presented in Fig.~\ref{fig:gs}(c). 
For these lattice sizes, we find the fidelity per qutrit/site $f:=\bra{00}_L \rho \ket{00}_L^{1/n_{\textrm{sites}}}$, as discussed in Appendix~\ref{app:fidelity_bounds}, to be:
\begin{align*}
	f &\in [0.959(3), 0.978(2)], \quad\textrm{for } 6\times2, \\
	f &\in [0.972(2), 0.985(1)], \quad\textrm{for } 4\times4.
\end{align*}
We observe a slightly lower per-qutrit fidelity on the $6 \times 2$ lattice compared to the $4 \times 4$ lattice. This difference might be attributed to the constrained geometry of the $6 \times 2$ lattice. This forces the unitary preparation protocol to order the plaquette preparation in a more chain-like manner, impacting circuit depth. 

Figure~\ref{fig:gs_smaller_sizes}(c) illustrates the effect on energy densities when heralded shots are not discarded. While a slight decrease in energy densities is observed, the overall impact is relatively small. 

To evaluate the fidelity of the prepared state independently of the quality of qubit measurements, it is crucial to consider the expectation values of stabilizer projectors after employing measurement error mitigation. State preparation and measurement (SPAM) error mitigation accounts for state preparation and readout errors. In practice, since measurement errors significantly outweigh state preparation errors, we implement SPAM error mitigation in a simplified form by constructing the measurement error transition matrix based on Quantinuum's H2-1 prior device characterization parameters. Specifically, we use $p(\textrm{measure 0} | \textrm{qubit is 1}) = 2.37e-3$ and $p( \textrm{measure 1} | \textrm{qubit is 0}) = 0.82e-3$. The inverse transition matrix is then applied to each qubit by writing the raw probability distribution of measurement outcomes as a matrix product state of bond dimension $n_{\textrm{shots}}$. 

In Fig.~\ref{fig:gs_smaller_sizes}d we show the data while employing only SPAM error mitigation. In contrast, Fig.~\ref{fig:gs_smaller_sizes}e displays the values where, in addition to SPAM error mitigation we discard heralded shots based on qutrit measuments. We find that both approaches, independently, contribute to an improved fidelity of the prepared state. 

\begin{figure*}[!ht]
\centering    \includegraphics[width=1.00\linewidth]{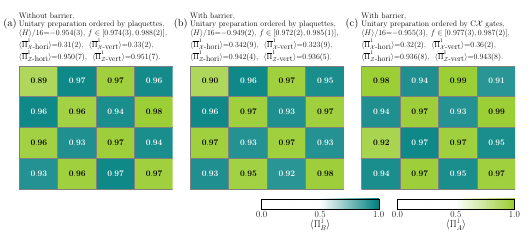}
\caption{\textbf{Barrier insertion and gate ordering at the control-$\mathcal{X}$ level for state preparation.} (a) Presents preparation data for the ground state without a barrier before inal measurements of qutrits. (b) This shows preparation data with a barrier, identical to Fig.~\ref{fig:gs_smaller_sizes}b and included for completeness. (c) We demonstrate preparation data optimized for circuit depth at the C-$\mathcal{X}$ gate level, rather than at the plaquette level. 
\label{fig:gs_4x4_barrier_comparisons}}
\end{figure*}

Since the depth of the unitary state preparation circuit is not uniform across all qutrits/qubits, we introduce a barrier at the end of the circuit before final measurements. This ensures that the system reaches the full $\mathbb{Z}_3$ ground state wavefunction before measurements collapse it into a product state. In Fig.~\ref{fig:gs_4x4_barrier_comparisons}(a,b), we contrast the impact of barrier insertion on the state fidelity. 
While there is a slight decrease in quality, the overall impact is negligible. Furthermore, by optimizing the state preparation for circuit depth, we can slightly improve the overall fidelity (cf. Fig.~\ref{fig:gs_4x4_barrier_comparisons}c).This optimization involves parallelizing gate operations not at the plaquette level but at the individual control-$\mathcal{X}$ gate level. 

\begin{figure*}[t]
	\includegraphics[width=0.99\linewidth]{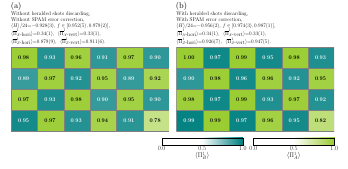}
	\caption{\textbf{Ground state energy densities for 6$\times$4 lattice} Expectation values for stabilizers $\Pi_A^1$ and $\Pi_B^1$. (a) We do not discard heralded shots where qutrits are measured outside the qutrit space, and SPAM error mitigation is applied. (b) We discard heralded shots and employ SPAM error mitigation.  \label{fig:6x4_comparisons}}
\end{figure*}

We observe similar behavior for the $6 \times 4$ lattice in terms of improving fidelities when discarding heralding shots and applying SPAM mitigation (see Fig.~\ref{fig:6x4_comparisons}).

\clearpage

\section{Topological qutrit on a 6$\times$2 Lattice}
\label{app:cc-entanglement-6x2}

This appendix presents a detailed investigation of entanglement transfer on a smaller $6\times2$ lattice. 
The reduced hardware costs associated with this smaller lattice allowed us to conduct a more comprehensive study compared to the larger $6\times4$ lattice. 

\begin{figure*}[!t]
	\centering    \includegraphics[width=0.85\linewidth]{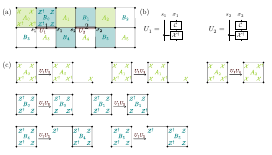}
	\caption{\textbf{CC defect pairs} Two CC defect pairs, labeled 1 and 2, are created on a $6\times2$ lattice. Defect pair 1 is generated by applying the unitary operator $U_1$ to qutrits $s_1$ and $\sigma_1$. Defect pair 2 is created by applying the unitary operator $U_2$ to qutrits $s_2$ and $\sigma_2$. The colored plaquettes highlight stabilizers that are non-trivially transformed under the action of $U_1U_2$. (b) The circuit constructions of the unitary operators $U_1$ and $U_2$ are shown. (c) Transformed stabilizers under the combined action of $U_1U_2$. The stabilizers $A_0$ and $B_4$ define the endpoints of defect pair 1, while stabilizers $A_1$ and $B_5$ define the endpoints of defect pair 2. 
		\label{fig:6x2_entangle_defects_construction}}
\end{figure*}

\begin{figure*}[!t]
	\centering    \includegraphics[width=0.70\linewidth]{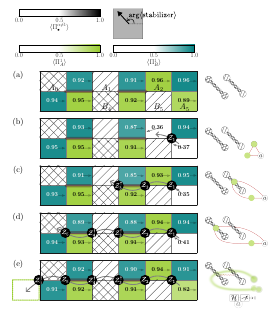}
	\caption{\textbf{Entanglement transfer to pairs of charge conjugation defects on a 6$\times$2 lattice} 
		We prepare the $\ZZ_3$ ground state and then create two pairs of charge conjugation defects. (a-e) The left-hand side displays data representing the expectation values of stabilizer projectors during different stages of entanglement transfer. The right-hand side provides a schematic representation to illustrate the corresponding state. Maximum error in estimating expectation value is $0.023$.
		\label{fig:6x2_entangle_defects}}
\end{figure*}

We begin by preparing the ground state and then apply the circuits $U_1$ and $U_2$ to create two pairs of defects (Fig.~\ref{fig:6x2_entangle_defects_construction}(a,b)). The definitions of the transformed stabilizers are given in Fig.~\ref{fig:6x2_entangle_defects_construction}c. The resulting expectation values of the stabilizers are shown in Fig.~\ref{fig:6x2_entangle_defects}a. We then initialize an ancilla qutrit `$a$' in the state $\frac{1}{\sqrt{3}}(\ket{0_a}+\ket{1_a}+\ket{2_a})$ and use it as a control to apply a control-$\mathcal{Z}_1$ gate. This entangles the stabilizers $A_2$ and $A_5$, resulting in the following state including the ancilla: $(\ket{0_a}\ket{\vac\vac}+\ket{1_a}\ket{e\ebar }+\ket{2_a}\ket{\ebar e})$. Fig.~\ref{fig:6x2_entangle_defects}b shows the resulting stabilizer values. The fact that stabilizers $A_2$ and $A_5$ are locally in a completely mixed state is demonstrated by their values, which are approximately $1/3$.

By further applying $\mathcal{Z}_2$ and $\mathcal{Z}_3^\dagger$ (cf. Fig.~\ref{fig:cc_construction}c), we effectively shift the position of one end of the charge pair. This alters the state of defect pair 2, resulting in a joint state of: $(\ket{0_a}\ket{\vac\vac}\ket{\vac_2} + \ket{1_a}\ket{e\ebar}\ket{e_2} + \ket{2_a}\ket{\ebar e}|{\ebar_2}\rangle)$. Next, we apply $\mathcal{Z}_4^\dagger$ and $\mathcal{Z}_5$ (as shown in Fig.~\ref{fig:cc_construction}d) to also entangle defect pair 1. By applying the operator $\mathcal{Z}_6^\dagger$, we fuse the charge anyon back with its partner as it completes a loop around the torus (Fig.~\ref{fig:6x2_entangle_defects_construction}e). This leaves the two defect pairs entangled with the ancilla, resulting in the following state:
\begin{align}
	\ket{0_a}|{\vac}_1{\vac}_2\rangle + \ket{1_a}|e_1\ebar_2\rangle + \ket{2_a}|\ebar_1e_2\rangle. \nonumber
\end{align}
The ancilla qutrit can be decoupled by measurement in the $\mathcal{X}$ basis. Depending on the measurement outcome, we are left with three distinct topological qutrit state and get the following values for the expectation values of stabilizers projectors:
\begin{center}
	\setlength{\tabcolsep}{8pt} 
	\renewcommand{\arraystretch}{1.7} 
	\begin{tabular}{c p{40mm} p{10mm} p{10mm} p{10mm} p{52mm} } 
		\hline 
		Ancilla outcome & State of defect pairs & ${\Pi^1_{A_0}}$ & ${\Pi^1_{A_1}}$ & ${\Pi^1_{A_0A_1}}$ & (${\Pi^1_{\Xent \Xent^\dagger}}$, ${\Pi^{\omega}_{\Xent\Xent^\dagger}}$, ${\Pi^{\bar \omega}_{\Xent \Xent^\dagger}}$) \\[.5ex] 
		\hline 
		0 & $\ket{\vac_1\vac_2} + |e_1\ebar_2\rangle + |\ebar_1e_2\rangle$ 
		& 0.31(3) \newline 0.31(3)$^*$& 0.33(3) \newline 0.33(3)$^*$ & 0.83(2) \newline 0.86(2)$^*$
		& (0.92(2), 0.05(1), 0.03(1)) \newline (0.94(2), 0.04(1), 0.017(8))$^*$ \\ 
		
		\hline 
		1 & $\ket{\vac_1 \vac_2} + \omega |e_1\ebar_2\rangle + \bar \omega |\ebar_1e_2\rangle$ 
		& 0.31(3) \newline 0.31(3)$^*$& 0.27(3) \newline 0.27(3)$^*$ & 0.89(2) \newline 0.92(2)$^*$
		& (0.017(9), 0.94(2), 0.04(1)) \newline (0.008(6), 0.96(1), 0.03(1))$^*$\\

		\hline 
		2 & $\ket{\vac_1 \vac_2} + \bar \omega |e_1\ebar_2\rangle + \omega |\ebar_1e_2\rangle$ 
		& 0.31(3) \newline 0.31(3)$^*$ & 0.32(3) \newline 0.32(3)$^*$ & 0.82(3) \newline 0.84(2)$^*$
		& (0.06(2), 0.07(2), 0.87(2)) \newline (0.05(1), 0.06(2), 0.89(2))$^*$\\ 
		[0.5ex] 
		\hline
	\end{tabular}
\end{center}
Numbers marked with an asterisk (*) were calculated using SPAM error mitigation, as outlined in Appendix~\ref{app:gs_data}. 
The internal states of defect pairs are characterized by $A_0$ and $A_1$. While the projectors $\Pi^1_{A_0}$ and $\Pi^1_{A_1}$ have values close to 1/3, indicating a locally mixed state, the joint projector $\Pi^1_{A_0A_1}$ exhibits a high value for each ancilla outcome, suggesting a globally entangled state. The action of the gray string operator (illustrated in Figure~\ref{fig:6x2_entangle_defects}e, with $\mathcal{Z}_1 \mathcal{Z}_2 \mathcal{Z}_3^\dagger \mathcal{Z}_4^\dagger \mathcal{Z}_5 \mathcal{Z}_6^\dagger$)
has the same effect as applying the logical $\Xent \Xent^\dagger$ stabilizer to the internal states of the topological qutrit, as prescribed in \eqref{eq:x_ent}. $\Xent \Xent^\dagger$ attains its maximum value in distinct sectors for each qutrit state, enabling their identification. 

\clearpage

\section{Defects in the $\ZZ_3$ TC}
\label{app:z3_defects}
The automorphism group of the $\ZZ_3$ TC is $\ZZ_2^2$, generated by two electromagnetic dualities $e\leftrightarrow m$ and $e\leftrightarrow \bar m$. Their product is charge conjugation $e\leftrightarrow \bar e$ and $m \leftrightarrow \bar m$. The quantum dimension of the em duality defects are $\sqrt 3$ and charge conjugation is $3$.

\subsection{Electro-Magnetic Duality Defects}

In the rotated toric code, we can create $em$-duality defects by measuring $\mathcal{Y} = \mathcal{X} \mathcal{Z}$ along a line and then applying some feed-forward correction based on the measurement outcomes. 

Measuring $\mathcal{Y}$ on a single vertex fuses the $\mathcal{X}$ and $\mathcal{Z}$ stabilizers with support on that site together in the following way:

\begin{equation}
\raisebox{-40pt}{
\begin{tikzpicture}
    \node at (0, 0) {
        \begin{tikzpicture}
            \draw[thick, fill=blue!20] (-3.5, 0) -- (-2.5, 0) -- (-2.5, 1) -- (-3.5, 1) -- (-3.5, 0);
            
            \node [fill=white,inner sep=1pt] at (-3.5,1) {$\mathcal{Z^{\dagger}}$};
            \node [fill=white,inner sep=1pt] at (-2.5,1) {$\mathcal{Z}$};
            \node [fill=white,inner sep=1pt] at (-3.5,0) {$\mathcal{Z}^{\dagger}$};
            \node [fill=white,inner sep=1pt] at (-2.5,0) {$\mathcal{Z}$};
        \end{tikzpicture}
    };
    \node at (0, -2) {
        \begin{tikzpicture}
            \draw[thick, fill=green!20] (-3.5, 0) -- (-2.5, 0) -- (-2.5, 1) -- (-3.5, 1) -- (-3.5, 0);
            
            \node [fill=white,inner sep=1pt] at (-3.5,1) {$\mathcal{X}$};
            \node [fill=white,inner sep=1pt] at (-2.5,1) {$\mathcal{X}$};
            \node [fill=white,inner sep=1pt] at (-3.5,0) {$\mathcal{X}^{\dagger}$};
            \node [fill=white,inner sep=1pt] at (-2.5,0) {$\mathcal{X}^{\dagger}$};
        \end{tikzpicture}
    };
    \node at (2, 0) {
        \begin{tikzpicture}
            \draw[thick, fill=green!20] (-3.5, 0) -- (-2.5, 0) -- (-2.5, 1) -- (-3.5, 1) -- (-3.5, 0);
            
            \node [fill=white,inner sep=1pt] at (-3.5,1) {$\mathcal{X}$};
            \node [fill=white,inner sep=1pt] at (-2.5,1) {$\mathcal{X}$};
            \node [fill=white,inner sep=1pt] at (-3.5,0) {$\mathcal{X}^{\dagger}$};
            \node [fill=white,inner sep=1pt] at (-2.5,0) {$\mathcal{X}^{\dagger}$};
        \end{tikzpicture}
    };
    \node at (2, -2) {
        \begin{tikzpicture}
            \draw[thick, fill=blue!20] (-3.5, 0) -- (-2.5, 0) -- (-2.5, 1) -- (-3.5, 1) -- (-3.5, 0);
            
            \node [fill=white,inner sep=1pt] at (-3.5,1) {$\mathcal{Z^{\dagger}}$};
            \node [fill=white,inner sep=1pt] at (-2.5,1) {$\mathcal{Z}$};
            \node [fill=white,inner sep=1pt] at (-3.5,0) {$\mathcal{Z}^{\dagger}$};
            \node [fill=white,inner sep=1pt] at (-2.5,0) {$\mathcal{Z}$};
        \end{tikzpicture}
    };
    \draw[->] (3.5,-1) -- node[above,midway] {measure $\mathcal{Y}$} (5,-1);
    \node at (6.5, -1) {
    \begin{tikzpicture}
        \draw[thick, fill=yellow!20] (-3.5, 0) -- (-2.5, 0) -- (-2.5, 2) -- (-3.5, 2) -- (-3.5, 0);  
        \node [fill=white,inner sep=1pt] at (-3.5,2) {$\mathcal{Z}^\dagger$};
        \node [fill=white,inner sep=1pt] at (-2.5,2) {$\mathcal{Z}$};
        \node [fill=white,inner sep=1pt] at (-3.5,1) {$\mathcal{X}\mathcal{Z^{\dagger}}$};
        \node [fill=white,inner sep=1pt] at (-3.5,0) {$\mathcal{X}^{\dagger}$};
        \node [fill=white,inner sep=1pt] at (-2.5,0) {$\mathcal{X}^\dagger$};
        \filldraw[color=black, fill=red!60, thick] (-2.5, 1) circle (3pt);
    \end{tikzpicture}
    };
    \node at (9.5, -1) {
    \begin{tikzpicture}
         \draw[thick, fill=yellow!20] (-3.5, 0) -- (-2.5, 0) -- (-2.5, 2) -- (-3.5, 2) -- (-3.5, 0);
        \node [fill=white,inner sep=1pt] at (-3.5,2) {$\mathcal{X}^\dagger$};
        \node [fill=white,inner sep=1pt] at (-2.5,2) {$\mathcal{X}^\dagger$};
        \node [fill=white,inner sep=1pt] at (-3.5,0) {$\mathcal{Z}$};
        \node [fill=white,inner sep=1pt] at (-2.5,0) {$\mathcal{Z}^\dagger$};
        \node [fill=white,inner sep=1pt] at (-2.5,1) {$\mathcal{XZ}^\dagger$};
        \filldraw[color=black, fill=red!60, thick] (-3.5, 1) circle (3pt);
    \end{tikzpicture}
    };
    \filldraw[color=black, fill=red!60, thick] (8, -1) circle (3pt);
    \node at (8, -0.5) {$\mathcal{X}\mathcal{Z}$};
\end{tikzpicture}
}
\end{equation}
Additionally, there is a ``nonlocal'' stabilizer spanning both defect endpoints:
\begin{center}
    \begin{tikzpicture}
        \draw[thick, fill=yellow!20] (-3.5, 0) -- (-1.5, 0) -- (-1.5, 1) -- (-3.5, 1) -- (-3.5, 0);
        \node [fill=white,inner sep=1pt] at (-3.5,1) {$\mathcal{Z}^\dagger$};
        \node [fill=white,inner sep=1pt] at (-1.5,1) {$\mathcal{X}^\dagger$};
        \node [fill=white,inner sep=1pt] at (-3.5,0) {$\mathcal{Z}^\dagger$};
        \node [fill=white,inner sep=1pt] at (-1.5,0) {$\mathcal{X}$};
        \node [fill=white,inner sep=1pt] at (-2.5,1) {$\mathcal{X}^\dagger \mathcal{Z}$};
        \filldraw[color=black, fill=red!60, thick] (-2.5, 0) circle (3pt);
    \end{tikzpicture}
\end{center}
In this example, since we have created the minimum length defect line, this nonlocal stabilizer is the same weight as the endpoint stabilizers. As we grow the defect line, however, this defect will grow as well, so that it always connects the two endpoints.

Using these stabilizers, we can verify that crossing the defect line (here, it consists of the single measured vertex) transmutes an $e$ to an $m$ anyon. Consider a $e$ anyon passing through the defect from top to bottom; in order to satisfy both defect stabilizers, the string must change from $\mathcal{Z}$ to $\mathcal{X}^\dagger$, meaning it becomes an $m$ anyon on the other side. To keep the overall state neutral, the defect must have absorbed an $e \bar{m}$ anyon---this is captured by the fact the nonlocal stabilizer now has an eigenvalue of $\omega^\dagger$ rather than $1$. 
\

\subsection{Charge Conjugation Defects}

We present here unitary and measurement-based circuits for creating such an open charge conjugation line. A more detailed discussion can be found in \cite{lyons-2024}, in addition to a general derivation of anyon and defect ribbons for generic quantum doubles. 

\subsubsection{Explicit Circuits}

First, we give an example circuit for creating a pair of charge conjugation defects. The shaded plaquettes indicate the location of the defects created by this circuit.

\begin{equation}
\raisebox{-60pt}{
    \begin{tikzpicture}
        \node at (0, 0){
            \begin{tikzpicture}[scale=0.4]
                \draw[fill=black!40] (0, -2) -- (-2, -2) -- (-2, 0) -- (0, 0) -- (0, -2);
                \draw[fill=black!40] (6, 6) -- (6, 8) -- (4, 8) -- (4,6) -- (6, 6);
                \draw[thick] (0, 0) -- (6, 0) -- (6, 6) -- (0, 6) -- (0, 0);
                \draw[thick] (2, 0) -- (2, 6);
                \draw[thick] (4, 0) -- (4, 6);
                \draw[thick] (0, 2) -- (6, 2);
                \draw[thick] (0, 4) -- (6, 4);
                \filldraw[red] (0, 0) circle (6pt);
                \filldraw[red] (2, 2) circle (6pt);
                \filldraw[red] (4, 4) circle (6pt);
                \node[red] at (0.75, 0.5) {$a_1$};               \node[red] at (2.75, 2.5) {$a_2$}; 
                \node[red] at (4.75, 4.5) {$a_3$};

                \filldraw[blue] (0, 2) circle (6pt);
                \filldraw[blue] (2, 4) circle (6pt);
                \filldraw[blue] (4, 6) circle (6pt);
                \node[blue] at (-0.75, 2.5) {$\alpha_1$};               \node[blue] at (1.25, 4.5) {$\alpha_2$}; 
                \node[blue] at (3.25, 6.5) {$\alpha_3$}; 
                
            \end{tikzpicture}
        };
        \node at (5, 0) {
            \begin{tikzpicture}
                \begin{yquant}[vertical]
                    [red]
                    qubit {$\ket{a_1}$} a[1];
                    [blue]
                    qubit {$\ket{\alpha_1}$} alp[1];
                    [red]
                    qubit {$\ket{a_2}$} a[+1];
                    [blue]
                    qubit {$\ket{\alpha_2}$} alp[+1];
                    [red]
                    qubit {$\ket{a_3}$} a[+1];
                    [blue]
                    qubit {$\ket{\alpha_3}$} alp[+1];
                    
                    box {$\mathcal{X}$} a[1] | a[0];
                    box {$\mathcal{X}$} a[2] | a[1];
                    
                    barrier (a, alp);
                    
                    align alp;

                    box {$\mathcal{X}$} alp[0] | a[0];
                    box {$\mathcal{X}$} alp[1] | a[1];
                    box {$\mathcal{X}$} alp[2] | a[2];
                    
                    barrier (a, alp);
                    
                    box {$\mathcal{X}^{\dagger}$} a[2] | a[1];
                    box {$\mathcal{X}^{\dagger}$} a[1] | a[0];
                    
                \end{yquant}
            \end{tikzpicture}
        };
    \end{tikzpicture}
}
\label{eq:explicit-cc-circuit}
\end{equation}

We go through the derivation of this circuit below.

\subsubsection{Derivation}

Consider the $\mathbb{Z}_3$ toric code with the following stabilizers (on alternating plaquettes, as in the main text):

\begin{center}
    \begin{tikzpicture}
        \draw[thick, fill=blue!20] (-3.5, 0) -- (-2.5, 0) -- (-2.5, 1) -- (-3.5, 1) -- (-3.5, 0);
        
        \node [fill=white,inner sep=1pt] at (-3.5,1) {$\mathcal{Z^{\dagger}}$};
        \node [fill=white,inner sep=1pt] at (-2.5,1) {$\mathcal{Z}$};
        \node [fill=white,inner sep=1pt] at (-3.5,0) {$\mathcal{Z}^{\dagger}$};
        \node [fill=white,inner sep=1pt] at (-2.5,0) {$\mathcal{Z}$};
    \end{tikzpicture}\qquad
    \begin{tikzpicture}
        \draw[thick, fill=green!20] (-3.5, 0) -- (-2.5, 0) -- (-2.5, 1) -- (-3.5, 1) -- (-3.5, 0);
        
        \node [fill=white,inner sep=1pt] at (-3.5,1) {$\mathcal{X}$};
        \node [fill=white,inner sep=1pt] at (-2.5,1) {$\mathcal{X}$};
        \node [fill=white,inner sep=1pt] at (-3.5,0) {$\mathcal{X}^{\dagger}$};
        \node [fill=white,inner sep=1pt] at (-2.5,0) {$\mathcal{X}^{\dagger}$};
    \end{tikzpicture}
\end{center}

\begin{figure}
    \centering
    \includegraphics[width=0.75\linewidth]{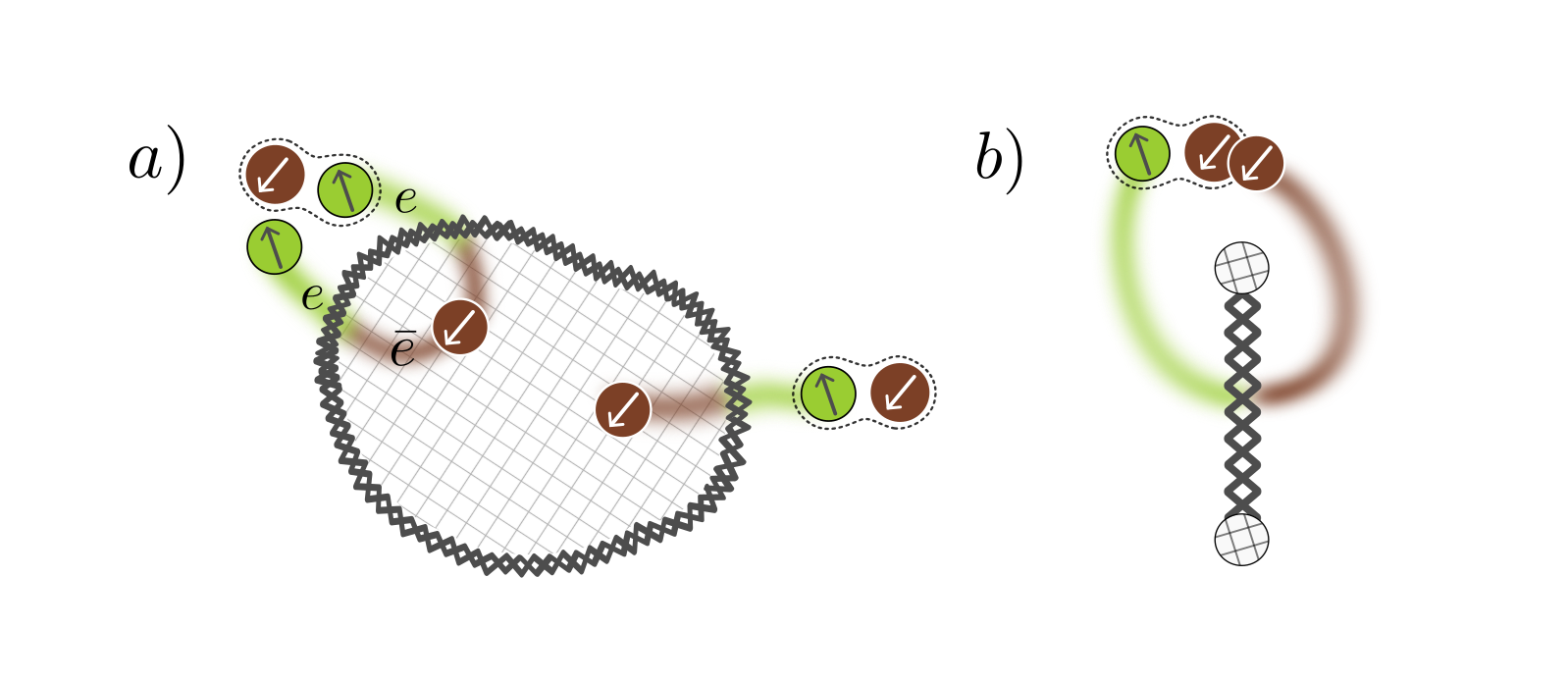}
    \caption{\textbf{Closed vs. open charge conjugation defect lines:} a) Applying $\prod \mathcal{C}$ in the shaded region creates a closed charge conjugation boundary; any anyon that crosses the boundary will be transmuted to its conjugate. Given the region is closed, any closed anyon loop must intersect the boundary an even number of times, meaning there are no measurable fusion results to determine the presence of such a boundary. b) An open charge conjugation has point defects living at its ends. Anyons can braid nontrivially with these defects, leading to nontrivial fusion outcomes.}
    \label{fig:cc-properties}
\end{figure}

Charge conjugation is a global symmetry of the $\mathbb{Z}_3$ toric code; acting with $\mathcal{C}$ on every degree of freedom preserves the ground state manifold and the braiding properties of any excitations. We can create a closed charge conjugation boundary by acting with charge conjugation in some finite region $A$ of the lattice (see Fig. \ref{fig:cc-properties}a). This doesn't affect the stabilizers that are completely contained in the region $A$ or the region $A^\perp$---only the stabilizers along the \emph{boundary} $\partial A$ will get modified:

\begin{center}
    \begin{tikzpicture}
        \draw[thick, fill=blue!20] (-3.5, 0) -- (-2.5, 0) -- (-2.5, 1) -- (-3.5, 1) -- (-3.5, 0);
        
        \node [fill=white,inner sep=1pt] at (-3.5,1) {$\mathcal{Z^{\dagger}}$};
        \node [fill=white,inner sep=1pt] at (-2.5,1) {$\mathcal{Z}^{\dagger}$};
        \node [fill=white,inner sep=1pt] at (-3.5,0) {$\mathcal{Z}$};
        \node [fill=white,inner sep=1pt] at (-2.5,0) {$\mathcal{Z}^{\dagger}$};
    \end{tikzpicture}\qquad
    \begin{tikzpicture}
        \draw[thick, fill=green!20] (-3.5, 0) -- (-2.5, 0) -- (-2.5, 1) -- (-3.5, 1) -- (-3.5, 0);
        
        \node [fill=white,inner sep=1pt] at (-3.5,1) {$\mathcal{X}$};
        \node [fill=white,inner sep=1pt] at (-2.5,1) {$\mathcal{X}$};
        \node [fill=white,inner sep=1pt] at (-3.5,0) {$\mathcal{X}^{\dagger}$};
        \node [fill=white,inner sep=1pt] at (-2.5,0) {$\mathcal{X}$};
    \end{tikzpicture}
\end{center}

When an $e$ or $m$ anyon crosses this boundary, it will get mapped to its conjugate particle. We can see this by considering what strings are valid (i.e. only create excitations at their ends) crossing the boundary. Away from the boundary, the usual $e$-string ($\mathcal{Z}$ along a diagonal) commutes with the unmodified stabilizers. When crossing the boundary, we have to have to switch to an $\mathcal{Z}^{\dagger}$ since the $e$-string now overlaps with two $\mathcal{X}^{\dagger}$ operators on the boundary $X$ plaquette. In this way, the $e$ particle has been turned into an $\bar e$ (see Fig.~\ref{fig:cc-properties}a). 

In some sense, however, all we have done is change the anyon labeling conventions inside region $A$. There will be no physically measurable effects of the $e \rightarrow \bar{m}$ transformation, since if we bring this anyon to fuse back with its partner, it will cross the charge conjugation boundary a second time, returning to its original state (see Fig.~\ref{fig:cc-properties}a again). 

More nontrivially, we can consider creating an open charge conjugation line, with two charge conjugation \emph{point} defects at either end. These defects cannot be created by acting with $\mathcal{C}$ in a bulk region. If such a pair of defects are present in the system, we can actually end up with measurable effects. For instance, if we have an open charge conjugation line, we can braid an $e$ around one endpoint; since it crosses the defect line only once, when it reaches its $\bar e$ partner, they will fuse back to a single $e$ (see Fig.~\ref{fig:cc-properties}b). The residual $\bar e$ needed to preserve overall neutrality of the state is now stored nonlocally in the internal Hilbert space of the charge conjugation defect line. 

Our goal is to find a circuit capable of creating such open charge conjugation defect lines. Essentially, we want a circuit which has the same action as a bulk charge conjugation symmetry membrane on the $\mathbb{Z}_3$ toric code ground state, but which is localized to the boundary of the membrane. This boundary circuit can be straightforwardly truncated to yield an creation operator for a pair of charge conjugation defects.

To derive and explain Eq.~\ref{eq:explicit-cc-circuit}, note that $\prod \mathcal{C}$ is a global $\mathbb{Z}_2$ symmetry of the $\mathbb{Z}_3$ paramagnet $\prod_{v} \left( \ket{0} + \ket{1} + \ket{2} \right)$, along with the $\mathbb{Z}_3$ symmetry $\prod \mathcal{X}$. After applying the $\mathbb{Z}_3$ Kramers-Wannier map \cite{tantivasadakarn_hierarchy_2023}, the $\mathbb{Z}_3$ paramagnet is mapped to the $\mathbb{Z}_3$ toric code---we say that these two phases are dual to each other. Charge conjugation acts trivially in a finite region of the $\mathbb{Z}_3$ paramagnet---therefore, all the key properties of the charge conjugation symmetry membrane on the $\mathbb{Z}_3$ toric code are captured by the gauging map itself. By `pushing' the membrane from the $\mathbb{Z}_3$ toric code through the $\mathbb{Z}_3$ gauging map back to the paramagnet, we can simplify the bulk operator to some action purely at the membrane boundary.

The $\mathbb{Z}_3$ gauging map $\mathsf{KW}_{\mathbb{Z}_3}$ takes paramagnetic degrees of freedom (green plaquettes) to the vertex degrees of freedom in the following way:
\begin{equation}
\raisebox{-17pt}{
\begin{tikzpicture}
\node at (-0.5,0){
    \begin{tikzpicture}[scale=0.8]
    \draw[-, thick, fill=green!20] (-0.6,-0.6) -- (-0.6,0.6) -- (0.6, 0.6) -- (0.6, -0.6) -- (-0.6, -0.6);
    \node[inner sep=1pt] at (0,0) {$\mathcal X$};
    \end{tikzpicture}
};
\draw[->] (0.25,0) -- node[above,midway] {\footnotesize $\mathsf{KW}_{\mathbb{Z}_3}$} (1.25,0);
\node at (2,0){
    \begin{tikzpicture}[scale=0.8]
    \draw[-,thick, fill=green!20] (-0.6,-0.6) -- (-0.6,0.6) -- (0.6, 0.6) -- (0.6, -0.6) -- (-0.6, -0.6);
    \node[fill=white,inner sep=1pt] at (-0.6,0.6) {$\mathcal X$};
    \node[fill=white,inner sep=1pt] at (0.6,0.6) {$\mathcal X$};
    \node[fill=white,inner sep=1pt] at (-0.6,-0.6) {$\mathcal X^\dagger$};
    \node[fill=white,inner sep=1pt] at (0.6,-0.6) {$\mathcal X^\dagger$};
    \end{tikzpicture}
};
\node at (5.5,0){
    \begin{tikzpicture}[scale=0.8]
    \draw[-,thick, fill=green!20] (-0.6,-0.6) -- (-0.6,0.6) -- (0.6, 0.6) -- (0.6, -0.6) -- (-0.6, -0.6);
    \draw[-,thick, fill=green!20] (0.6,0.6) -- (0.6,1.8) -- (1.8, 1.8) -- (1.8, 0.6) -- (0.6, 0.6);
    \node[inner sep=1pt] at (0,0) {$\mathcal Z$};
    \node[inner sep=1pt] at (1.2,1.2) {$\mathcal Z^\dagger$};
    \end{tikzpicture}
};
\draw[->] (6.75,0) -- node[above,midway] {\footnotesize $\mathsf{KW}_{\mathbb{Z}_3}$} (7.75,0);
\node at (9,0){
    \begin{tikzpicture}[scale=0.8]
    \draw[-,thick, fill=green!20] (-0.6,-0.6) -- (-0.6,0.6) -- (0.6, 0.6) -- (0.6, -0.6) -- (-0.6, -0.6);
    \draw[-,thick, fill=green!20] (0.6,0.6) -- (0.6,1.8) -- (1.8, 1.8) -- (1.8, 0.6) -- (0.6, 0.6);
    \node[fill=white,inner sep=1pt] at (0.6,0.6) {$\mathcal Z$};
    \end{tikzpicture}
};
\end{tikzpicture}
}
\label{eq:z3-gauging-props}
\end{equation}
These expressions fully specify the action of the gauging map on $\mathcal{C}$, as
\begin{equation}
    \mathcal{C} = \ket{0}\bra{0} + \mathcal{X}\ket{1}\bra{1} + \mathcal{X}^\dagger\ket{2}\bra{2}  =\sum_{n=0}^2 \mathcal X^n T_{n}
\label{eq:cc-decomp}
\end{equation}
where $T_n = \ket{n}\bra{n} =\frac{1}{3}(1+ \omega^{-n}\mathcal Z +\omega^n\mathcal Z^\dagger)$. Using this decomposition, we can apply the gauging transformation rules to find the following identities:
\begin{equation}
\raisebox{-50pt}{
\begin{tikzpicture}
\node at (0, 0) {
    \begin{tikzpicture}
        \node at (-1.1, 0) { \footnotesize $\mathsf{KW}_{\mathbb{Z}_3} ~\cdot$};
        \node at (0,0){
            \begin{tikzpicture}[scale=0.75]
                \draw[-, thick, fill=green!20] (-0.6,-0.6) -- (-0.6,0.6) -- (0.6, 0.6) -- (0.6, -0.6) -- (-0.6, -0.6);
                \node[inner sep=1pt] at (0,0) {$\mathcal C$};
            \end{tikzpicture}
        };
        \node at (1.1, 0) {$= \sum\limits_{n=0}^2 $}; 
        \node at (1.9, 0) {\huge $($};
        \node at (2.75,0){
            \begin{tikzpicture}[scale=0.75]
            \draw[-, thick, fill=green!20] (-0.6,-0.6) -- (-0.6,0.6) -- (0.6, 0.6) -- (0.6, -0.6) -- (-0.6, -0.6);
            \node[fill=white,inner sep=1pt] at (0.6,0.7) {$\mathcal{X}$};
            \node[fill=white,inner sep=1pt] at (-0.6,0.6) {$\mathcal{X}$};
            \node[fill=white,inner sep=1pt] at (-0.6,-0.6) {$\mathcal{X}^{\dagger}$};
            \node[fill=white,inner sep=1pt] at (0.7,-0.6) {$\mathcal{X}^{\dagger}$};
            \end{tikzpicture}
        };
        \node at (3.6, 0) {\huge $)$};
        \node at (3.7, 0.4) {$n$};
        \node at (4.6, 0) {\footnotesize $\cdot ~ \mathsf{KW}_{\mathbb{Z}_3} ~\cdot$};
        \node at (5.9, 0) {
            \begin{tikzpicture}[scale=0.75]
                \draw[-, thick, fill=green!20] (-0.6,-0.6) -- (-0.6,0.6) -- (0.6, 0.6) -- (0.6, -0.6) -- (-0.6, -0.6);
                \node[inner sep=1pt] at (0,0) {$T_n$};
            \end{tikzpicture}
        };
    \end{tikzpicture}
}; 
\node at (0, -2) {
    \begin{tikzpicture}
        \node at (0,0){
            \begin{tikzpicture}[scale=0.8]
                \draw[-,thick, fill=green!20] (-0.6,-0.6) -- (-0.6,0.6) -- (0.6, 0.6) -- (0.6, -0.6) -- (-0.6, -0.6);
                \draw[-,thick, fill=green!20] (0.6,0.6) -- (0.6,1.8) -- (1.8, 1.8) -- (1.8, 0.6) -- (0.6, 0.6);
                \node[fill=white,inner sep=1pt] at (0.6,0.6) {$\mathcal C$};
            \end{tikzpicture}
        };
        \node at (2, 0) {\footnotesize $\cdot~ \mathsf{KW}_{\mathbb{Z}_3} = \sum\limits_{n, n'}$};
        \node at (4.25,0){
            \begin{tikzpicture}[scale=0.8]
                \draw[-,thick, fill=green!20] (-0.6,-0.6) -- (-0.6,0.6) -- (0.6, 0.6) -- (0.6, -0.6) -- (-0.6, -0.6);
                \draw[-,thick, fill=green!20] (0.6,0.6) -- (0.6,1.8) -- (1.8, 1.8) -- (1.8, 0.6) -- (0.6, 0.6);
                \node[fill=white,inner sep=1pt] at (0.6,0.6) {$\mathcal{X}^{n-n'}$};
            \end{tikzpicture}
        };
        \node at (6, 0) {\footnotesize $\cdot~ \mathsf{KW}_{\mathbb{Z}_3} ~\cdot$};
        \node at (7.75, 0) {
            \begin{tikzpicture}[scale=0.8]
                \draw[-,thick, fill=green!20] (-0.6,-0.6) -- (-0.6,0.6) -- (0.6, 0.6) -- (0.6, -0.6) -- (-0.6, -0.6);
                \draw[-,thick, fill=green!20] (0.6,0.6) -- (0.6,1.8) -- (1.8, 1.8) -- (1.8, 0.6) -- (0.6, 0.6);
                \node[inner sep=1pt] at (0,0) {$T_n$};
                \node[inner sep=1pt] at (1.2,1.2) {$T_n'$};
            \end{tikzpicture}
        };
    \end{tikzpicture}
};
\end{tikzpicture}
}
\end{equation}

\begin{figure}
    \centering
    \includegraphics[width=0.75\linewidth]{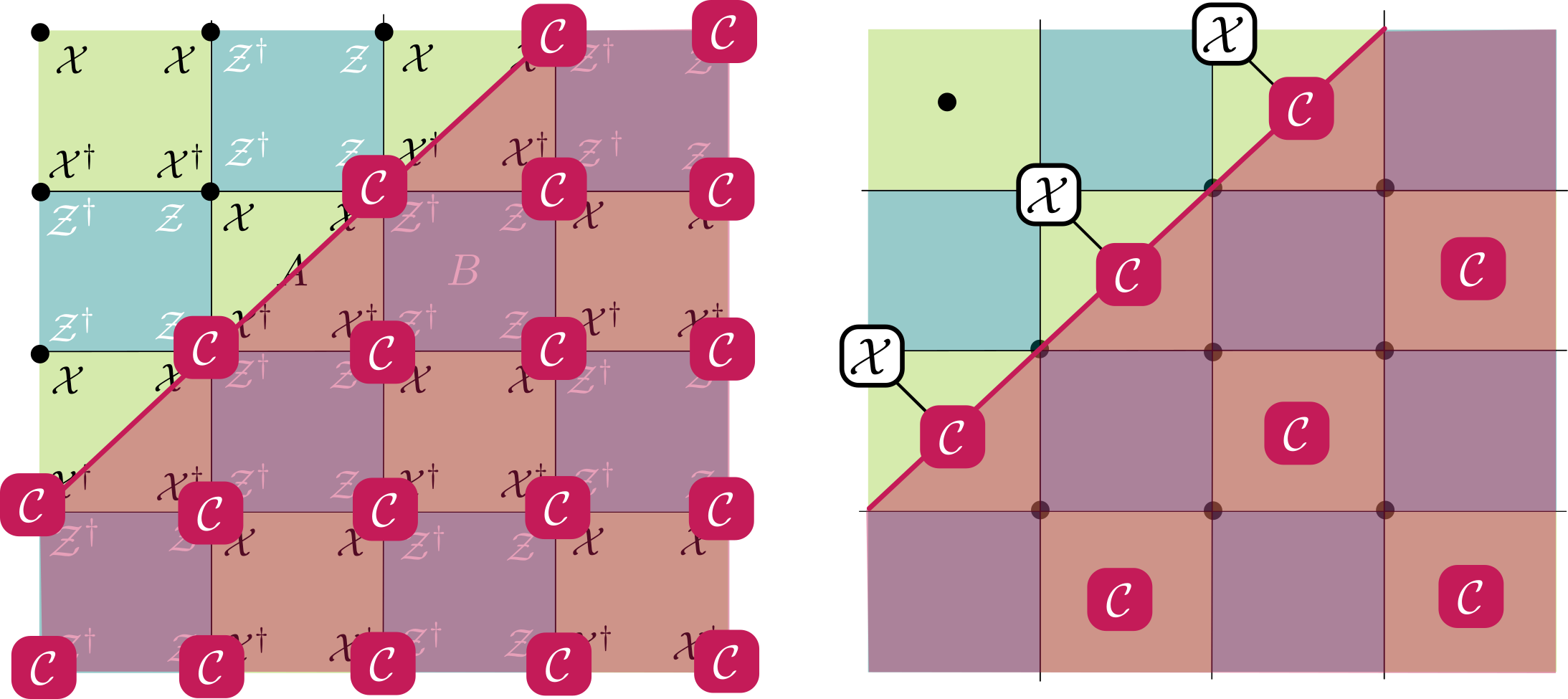}
    \caption{\textbf{A bulk charge conjugation membrane is equivalent to some boundary action:} Left: A bulk charge conjugation symmetry membrane acting on the toric code degrees of freedom, which creates a charge conjugation defect boundary. Right: Via the properties of the gauging map, the membrane is equivalent to a charge conjugation membrane on the dual paramagnetic degrees of freedom, plus a finite-depth circuit of control gates at the boundary, with the paramagnetic degrees of freedom as controls and the toric code degrees of freedom as targets. The charge conjugation gates act trivially on the paramagnet and can be ignored.}
    \label{fig:cc-derivation}
\end{figure}

We see that a single-site charge conjugation operator prior to the gauging map is equivalent to vertex operator after the gauging map, controlled on the pre-gauged state. Similarly, conjugation acting from the toric code side is equivalent to a product of $\text{C}\mathcal{X}$, controlled on the neighboring pre-gauged sites. Figure \ref{fig:cc-derivation} illustrates how the above identities can be used to reduce a bulk charge conjugation defect acting on the $\mathbb{Z}_3$ toric code to a boundary circuit plus a bulk charge conjugation action on the trivial paramagnet. However, the paramagnet is invariant under even a finite charge conjugation membrane, and so these gates can be ignored. Only the boundary circuit remains, given by: 
\begin{equation}
\prod\limits_{s \in \gamma} \text{C}\mathcal{X}^{\eta(v_s)}_{s \rightarrow v_s}
\end{equation}
where $\gamma$ is the desired charge conjugation boundary, $s$ is a paramagnetic degree of freedom lying on the boundary $\gamma$, $v_s$ is the neighboring toric code degree of freedom just outside the boundary (see Fig.~\ref{fig:cc-derivation}), and $\eta(e) = +1 (-1)$ if $v_s$ lies on the left (right) side of the plaquette corresponding to $s$.

We have partially derived the example unitary circuit. However, the above finite-depth circuit acts on both pre-gauged sites and gauged vertices; in the experiment, we only have access to the gauged toric code degrees of freedom. To circumvent this issue, we use the one-dimensional counterpart of the $\mathbb{Z}_3$ gauging map to access the necessary pre-gauged sites. This is valid due to the presence of a 1-form symmetry in the $\mathbb{Z}_3$ toric code; this 1-form symmetry acts like a global symmetry along a chosen 1D path, making the application of the 1D gauging map valid. The 1D gauging map essentially uncovers the pre-gauged degrees of freedom needed to act as controls for the finite-depth portion of the circuit. 

The 1D gauging map can be implemented unitarily via a linear-depth, sequential circuit: \begin{equation}
    \mathsf{KW}^{\text{1D};\gamma}_{\mathbb{Z}_3} = \prod_{i \in \gamma} C\mathcal{X}_{i, i+1}
\end{equation}

This completes the derivation of the charge conjugation defect circuit: (1) apply the 1D gauging map to reveal needed pre-gauged degrees of freedom, (2) use the derived finite-depth boundary circuit, (3) apply the (un)-gauging map to go back to toric code degreees of freedom everywhere. Written explicitly, we have: 
\begin{equation}
    U^\gamma_{c.c.} = \mathsf{KW}^{\text{1D};\gamma \dagger}_{\mathbb{Z}_3} \left ( \prod\limits_{s \in \gamma} \text{C}\mathcal{X}^{\eta(v_s)}_{s \rightarrow v_s} \cdot \mathcal{C}_{v_s} \right) \mathsf{KW}^{\text{1D};\gamma}_{\mathbb{Z}_3}
\end{equation}

\section{Coherently moving the end of a charge conjugation defect}\label{app:coherently_moving}

First, we define what coherently moving the end of a CC defect pair means operationally. Given a CC defect pair ending on site $A$ and site $B$, we have coherently moved the end of a CC defect pair at site $B$ if there is no remnant particle and the stabilizers at site $B$ return to their original form. Therefore, to coherently move one end of a CC defect from site $B$ to site $C$, we can apply another ribbon operator with ends at $B$ and $C$, and ensure that the two CC defects at site $B$ is in the vacuum fusion channel.

\begin{figure*}[t]
	\centering    \includegraphics[width=0.85\linewidth]{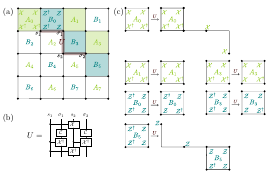}
	\caption{\textbf{Charge conjugation defect pair on a $4\times4$ lattice} (a) Square lattice on torus. A charge conjugation defect pair is created by a unitary action $U$ along the string of qutrits labeled $s_1$, $\sigma_1$, $s_2$, and $\sigma_2$. Stabilizers that transform non-trivially under the action of $U$ are highlighted by coloring the corresponding plaquettes. (b) Circuit used to create the charge conjugation defect pair. (c) Non-trivial transformations of $A_p$ and $B_p$ stabilizers under the unitary action. Stabilizers that remain invariant under unitary action are omitted. The stabilizers $A_0$ and $B_5$ define the endpoints of the defect pair and characterize its internal state.
	\label{fig:cc_construction}}
\end{figure*}

Next, we define the internal states of a CC defect. Consider the example of CC defect in Fig. \ref{fig:cc_construction}, upon applying the unitary $U$ (Fig. \ref{fig:cc_construction}b), the $A_0$ stabilizer at the start of the ribbon becomes non-local, and the $B_5$ stabilizer at the end of the ribbon becomes non-local (Fig. \ref{fig:cc_construction}c). The non-local $e$ and $m$ anyon determines the internal states of a CC defect, so we should label the internal state of a CC defect as $\ket{e^{a},m^{b}}$. If $a=0$ ($b=0)$, then it indicates that the non-local $e$ ($m$) stabilizer is not violated. This is a basis for the possible internal states of a CC defect pair, so the quantum dimension of one CC defect is $\sqrt{9}=3$, as expected.

Now, we discuss the reason why applying $U$ again is operationally equivalent to coherently moving the start of a CC defect. Initially, applying $U$ creates a CC defect pair in the internal state of $\ket{e^{0},m^{0}}$. The CC defect unitary circuit satisfy $U^2=1$; i.e. if we apply $U$ again, we would have coherently moved an end of a CC defect pair and returned to the vacuum. However, after crossing an $m$ anyon through the CC defect, the internal state becomes $\ket{e^{0},m}$. After applying another $U$, we are essentially overlaying another CC defect pair with internal state of $\ket{e^{0},m^{0}}$. At the start, the two ends fuse to vacuum, whereas at the end, the two ends does not fuse to vacuum, and hence we see a remnant particle of $m$.

\clearpage
\section{Supplementary Data and Figures}
\label{app:supplementary_data_figures}

\begin{figure*}[t]
\includegraphics[width=0.95\linewidth]{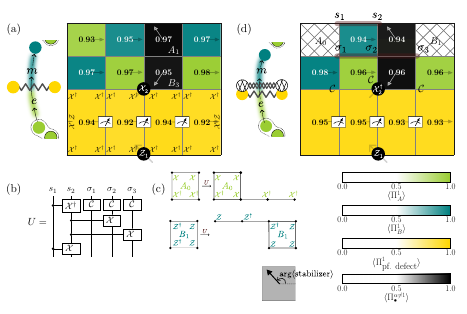}
    \caption{\textbf{Fusion of charge conjugation and parafermion defects} (a) We prepare the ground state and create a parafermion defect line by measuring three qutrits.  Depending on the measurement outcomes, we apply feed-forward operations to initialize the internal states of the parafermion defect line. We act with $\mathcal{Z}_1\mathcal{X}_2$ on qutrits on two sides of the defect line. This creates an $\ebar-m$ anyon pair on plaquettes $A_1-B_3$. (b) Circuit construction of a unitary to create a CC defect line. (c) Transformation of stabilizers at the endpoints of the CC defect line after applying the unitary $U$. (d) Starting with the ground state containing a parafermion defect line, we apply the unitary $U$ from (b) to create a CC defect line. We then deform the CC defect line by applying $\mathcal{C}$ gates. This effectively results in a conjugated parafermion line, as demonstrated by the fact that the correct string to create an $\ebar-m$ anyon is now $\mathcal{Z}_1\mathcal{X}_2^\dagger$. \label{fig:pf-conjugate}}
\end{figure*}

\begin{figure*}[t]
	\centering    \includegraphics[width=0.85\linewidth]{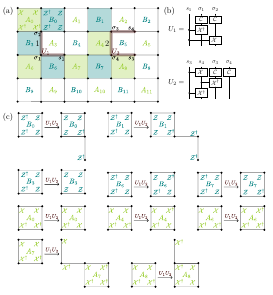}
	\caption{\textbf{Two charge conjugation defect pairs on a $6\times4$ lattice} (a) Two charge conjugation defect pairs, labeled 1 and 2, are created. Defect pair 1 is generated by applying the unitary $U_1$ to qutrits labeled $s_1$, $\sigma_1$, and $s_2$. Defect pair 2 is created by applying the unitary $U_2$ to qutrits labeled $s_3$, $s_4$, $\sigma_3$, and $\sigma_4$. Colored plaquettes indicate non-trivially transformed stabilizers. (b) Circuit construction of unitaries $U_1$ and $U_2$. (c) Transformation of stabilziers under the action $U_1U_2$. Stabilizers $B_0$ and $A_7$ define the endpoints of defect pair 1, while stabilizers $B_1$ and $A_8$ define the endpoints of defect pair 2. \label{fig:6x4_entangle_defects_construction}}
\end{figure*}

\clearpage

\end{document}